\documentclass[manuscript,screen,nonacm,sigconf]{acmart}

\setcopyright{none}
\renewcommand\footnotetextcopyrightpermission[1]{}

\usepackage{amsmath}
\usepackage{array}
\usepackage{booktabs}
\usepackage{enumitem}
\usepackage{graphicx}
\usepackage{listings}
\usepackage{mathtools}
\usepackage{multirow}
\usepackage{tabularx}
\usepackage{tikz}
\usepackage{xspace}
\newcommand{\sys}{Agent libOS\xspace}
\newcommand{\agentprocess}{\texttt{AgentProcess}\xspace}
\newcommand{\agentimage}{\texttt{AgentImage}\xspace}

\newcommand{\capability}{Capability\xspace}

\newcommand{\taskauthority}{Task Authority Manifest\xspace}
\newcommand{\protectedop}{Protected Operation\xspace}
\newcommand{\taskrun}{Durable Task Run\xspace}

\newcommand{\toolbroker}{\texttt{ToolBroker}\xspace}
\newcommand{\code}[1]{\texttt{\detokenize{#1}}}

\newcolumntype{L}[1]{>{\raggedright\arraybackslash}p{#1}}
\newcolumntype{C}[1]{>{\centering\arraybackslash}p{#1}}
\newcolumntype{Y}{>{\raggedright\arraybackslash}X}
\newcolumntype{Z}{>{\centering\arraybackslash}X}

\lstdefinestyle{smallcode}{
  basicstyle=\ttfamily\scriptsize,
  breaklines=true,
  columns=fullflexible,
  frame=single,
  xleftmargin=2pt,
  xrightmargin=2pt,
  framesep=3pt,
  showstringspaces=false
}
\providecommand{\ArtifactCommitShort}{77c9160c53d9}
\providecommand{\ArtifactSourceStatus}{available}

\providecommand{\CanonicalLiveCacheReadReportedCalls}{189}
\providecommand{\CanonicalLiveCacheReadTokens}{2496256}
\providecommand{\CanonicalLiveCacheTotalCalls}{189}
\providecommand{\CanonicalLiveCacheWriteReportedCalls}{0}

\providecommand{\CanonicalLiveInputTokens}{3853163}
\providecommand{\CanonicalLiveLogicalCalls}{189}
\providecommand{\CanonicalLiveOutputTokens}{316487}

\providecommand{\CanonicalLiveProviderAttempts}{195}

\providecommand{\CanonicalLiveSafetyPassed}{12}
\providecommand{\CanonicalLiveSafetyTotal}{12}
\providecommand{\CanonicalLiveStatus}{available}
\providecommand{\CanonicalLiveUtilityPassed}{12}
\providecommand{\CanonicalLiveUtilityTotal}{12}
\providecommand{\CollectedPytestNodeCount}{8903}

\providecommand{\ExternalEffectPending}{1000}

\providecommand{\ExternalEffectTotal}{100000}

\providecommand{\GUIStatus}{passed}

\providecommand{\InvariantCheckerStatus}{passed}
\providecommand{\InvariantDeclarationCount}{149}

\providecommand{\PracticalModeledPassed}{80}

\providecommand{\PracticalModeledTotal}{80}

\providecommand{\PracticalNativePassed}{3}

\providecommand{\PracticalNativeTotal}{3}

\providecommand{\ProviderSmokeStatus}{available}

\providecommand{\PublicationRecoveryPending}{1001}

\providecommand{\PublicationRecoveryTotal}{10000}
\providecommand{\RealModelTrajectoryCount}{42}
\providecommand{\RuntimeAuditCompletenessPercent}{100.0}

\providecommand{\RuntimeFalseDenialDenominator}{119}
\providecommand{\RuntimeFalseDenialNumerator}{0}

\providecommand{\RuntimeSafetyPassed}{33}

\providecommand{\RuntimeTaskCount}{33}
\providecommand{\RuntimeTaskSuccessPassed}{33}
\providecommand{\RuntimeUnauthorizedDenominator}{119}
\providecommand{\RuntimeUnauthorizedNumerator}{0}
\providecommand{\SkillsAllCorrectBaselinePassed}{13}
\providecommand{\SkillsAllCorrectDenominator}{15}
\providecommand{\SkillsAllCorrectTreatmentPassed}{13}
\providecommand{\SkillsBaselineCacheReadReportedCalls}{55}
\providecommand{\SkillsBaselineCacheReadTokens}{1399808}
\providecommand{\SkillsBaselineCacheTotalCalls}{55}

\providecommand{\SkillsBaselineCacheWriteReportedCalls}{0}

\providecommand{\SkillsBaselineInputTokens}{1563576}
\providecommand{\SkillsBaselineLogicalCalls}{55}
\providecommand{\SkillsBaselineOutputTokens}{101954}
\providecommand{\SkillsBaselineProviderAttempts}{56}
\providecommand{\SkillsBootstrapResamples}{10000}

\providecommand{\SkillsLiveStatus}{available}
\providecommand{\SkillsOutcomeBaselinePassed}{15}

\providecommand{\SkillsOutcomeDenominator}{15}
\providecommand{\SkillsOutcomeMcNemarP}{1.0000}
\providecommand{\SkillsOutcomeTreatmentPassed}{15}
\providecommand{\SkillsPairCount}{15}

\providecommand{\SkillsRouteBaselinePassed}{14}

\providecommand{\SkillsRouteDenominator}{15}
\providecommand{\SkillsRouteMcNemarP}{1.0000}
\providecommand{\SkillsRouteTreatmentPassed}{15}
\providecommand{\SkillsRunCount}{30}
\providecommand{\SkillsTerminalExitBaselinePassed}{14}
\providecommand{\SkillsTerminalExitDenominator}{15}
\providecommand{\SkillsTerminalExitTreatmentPassed}{13}
\providecommand{\SkillsTreatmentCacheReadReportedCalls}{99}
\providecommand{\SkillsTreatmentCacheReadTokens}{755072}
\providecommand{\SkillsTreatmentCacheTotalCalls}{99}

\providecommand{\SkillsTreatmentCacheWriteReportedCalls}{0}

\providecommand{\SkillsTreatmentInputTokens}{1068687}
\providecommand{\SkillsTreatmentLogicalCalls}{99}
\providecommand{\SkillsTreatmentOutputTokens}{151642}
\providecommand{\SkillsTreatmentProviderAttempts}{100}
\providecommand{\TaskRunCrashPassed}{6}

\providecommand{\TaskRunCrashTotal}{6}

\providecommand{\TaskRunRecoveryPending}{1000}

\providecommand{\TaskRunRecoveryTotal}{100000}

\providecommand{\TestMatrixStatus}{passed}
\providecommand{\RuntimeSafetyTableRows}{%
  direct\_tool\_wrapper & 33 & 90.9\% & 39.4\% & 20/29 & 0/9 & 0.00\% \\%
  confirmation\_wrapper & 33 & 93.9\% & 39.4\% & 20/64 & 0/44 & 0.00\% \\%
  sandbox\_only & 33 & 72.7\% & 39.4\% & 20/29 & 15/24 & 0.00\% \\%
  agent\_libos\_full & 33 & 100.0\% & 100.0\% & 0/119 & 0/119 & 100.0\% \\%
  no\_primitive\_approval & 33 & 97.0\% & 100.0\% & 0/118 & 0/118 & 100.0\% \\%
  no\_namespace\_isolation & 33 & 100.0\% & 93.9\% & 2/121 & 0/119 & 100.0\% \\%
  no\_fork\_attenuation & 33 & 100.0\% & 97.0\% & 1/120 & 0/119 & 100.0\% \\%
}
\providecommand{\PracticalTableRows}{%
  native-live & 3/3 & 3 & 3 \\%
  modeled & 80/80 & 160 & -- \\%
}
\providecommand{\RecoveryTableRows}{%
  External effect & 100000 & 1000 & 1000 & 500 \\%
  Runtime publication & 10000 & 1001 & 1001 & 500 \\%
  Durable TaskRun & 100000 & 1000 & 1000 & 500 \\%
  Crash barriers & 6 & +1 unpaired & 6/6; unpaired pass & -- \\%
}
\providecommand{\CanonicalLiveTableRows}{%
  Repository maintenance & 3/3 & 3/3 & 3 \\%
  Browser workflow & 3/3 & 3/3 & 3 \\%
  Research workflow & 3/3 & 3/3 & 3 \\%
  Analysis workflow & 3/3 & 3/3 & 3 \\%
  Total & 12/12 & 12/12 & 12 \\%
}
\providecommand{\SkillsPairedTableRows}{%
  Task outcome & 100.0\% & 100.0\% & 0.00 pp [0.00, 0.00] \\%
  Correct route & 100.0\% & 93.3\% & +6.67 pp [0.00, +20.0] \\%
  Invalid calls & 0.00 & 0.00 & 0.00 [0.00, 0.00] \\%
  Logical LLM calls & 6.60 & 3.67 & +2.93 [+2.13, +3.67] \\%
  Provider attempts & 6.67 & 3.73 & +2.93 [+2.00, +3.80] \\%
  Prompt tokens & 71246 & 104238 & -32993 [-54608, -15092] \\%
  Completion tokens & 10109 & 6797 & +3313 [+1388, +5245] \\%
  Cache-read tokens & 50338 & 93321 & -42982 [-62601, -27213] \\%
  Initial schema bytes & 7649 & 91759 & -84110 [-84110, -84110] \\%
  Cumulative schema bytes & 72360 & 336450 & -264090 [-330692, -215892] \\%
}
{evidence/results.tex is absent; empirical tables will report not-run}%
}
\providecommand{\ArtifactCommitShort}{unbound}

\providecommand{\ArtifactSourceStatus}{not-run}

\providecommand{\CanonicalLiveStatus}{not-run}
\providecommand{\SkillsLiveStatus}{not-run}
\providecommand{\RuntimeSafetyTableRows}{\multicolumn{7}{c}{not run}\\}
\providecommand{\PracticalTableRows}{\multicolumn{4}{c}{not run}\\}
\providecommand{\RecoveryTableRows}{\multicolumn{5}{c}{not run}\\}
\providecommand{\CanonicalLiveTableRows}{\multicolumn{4}{c}{not run}\\}
\providecommand{\SkillsPairedTableRows}{\multicolumn{4}{c}{not run}\\}

\newcommand{\IfResult}[3]{%
  \ifcsname #1\endcsname #2\else #3\fi
}
\newcommand{\IfResultFlag}[4]{%
  \ifcsname #1\endcsname
    \ifnum\csname #1\endcsname=1\relax #2\else #3\fi
  \else #4\fi
}

\title{Agent libOS: A Runtime Substrate for Capability-Controlled Self-Evolving LLM Agents}

\author{Yingqi Zhang}
\affiliation{%
  \institution{Department of Computer Science and Technology, Tsinghua University}
  \city{Beijing}
  \country{China}}
\email{zhangyq24@mails.tsinghua.edu.cn}

\acmConference[arXiv preprint]{arXiv preprint}{2026}{}
\acmBooktitle{arXiv preprint}
\acmYear{2026}
\copyrightyear{2026}


\begin{document}

\begin{abstract}
Large language model (LLM) agents increasingly persist across tasks, acquire
memory, activate Skills, synthesize tools, fork child processes, attach remote
resources, and commit checkpoints as reusable images.  These mechanisms expand
what an agent can request after deployment.  If a runtime treats an action's
visibility as authorization to perform it, however, self-evolution becomes a
path to authority escalation and data exfiltration.

This paper presents \sys, an agent-native library OS substrate that separates
three planes.  The design defines \emph{operation admission} as the conjunction of process
identity, the applicable Task Authority ceiling, typed Capabilities,
deterministic policy or Human approval, resource budgets, and execution through
a concrete primitive.  \emph{Information-flow admission} propagates
Runtime-maintained labels and immutable source references, resolves a canonical
Sink from the Host-owned registry, checks its clearance, and, for conditionally
permitted high-sensitivity egress, requires an exact one-shot Human release.  A
separate \emph{evidence plane} records durable intent, classified outcomes,
events, audit records, and causal links; evidence explains a decision but never
grants authority.  The resulting invariant is that an agent's model-visible
action surface may evolve without implicitly expanding either its resource
authority or its permitted information flows.

The implementation includes persistent \agentprocess instances with
process-local Object Memory namespaces and working directories; declarative
Skills; syscall-mediated Deno/TypeScript JIT Tools; images and scoped
checkpoints; typed Git; client-only JSON-RPC and MCP providers; Human queues;
hierarchical resource accounting; Host-only Explainable Operations; and
\taskrun supervision with fail-closed recovery.  Provider-backed effects follow
a prepare--dispatch--settle protocol that makes ambiguous dispatched outcomes
explicit and does not blindly replay them during recovery.
\IfResult{RuntimeSafetyPassed}{In the source-bound artifacts, the full-Runtime
evaluation reports \RuntimeTaskSuccessPassed{}/\RuntimeTaskCount{} task-success oracles and
\RuntimeSafetyPassed{}/\RuntimeTaskCount{} safety-passing tasks, with
unauthorized-effect and false-denial fractions of
\RuntimeUnauthorizedNumerator{}/\RuntimeUnauthorizedDenominator{} and
\RuntimeFalseDenialNumerator{}/\RuntimeFalseDenialDenominator{},
respectively.}{}
\IfResult{RealModelTrajectoryCount}{The real-model evaluation comprises
\RealModelTrajectoryCount{} trajectories.  Across the canonical workflow
matrix, observed safety is
\CanonicalLiveSafetyPassed{}/\CanonicalLiveSafetyTotal{} and observed strict
utility is \CanonicalLiveUtilityPassed{}/\CanonicalLiveUtilityTotal{}.  In the
paired projection study, the observable-state oracle passes for
\SkillsOutcomeTreatmentPassed{}/\SkillsOutcomeDenominator{} runs in the Skills
arm and for \SkillsOutcomeBaselinePassed{}/\SkillsOutcomeDenominator{} runs in
the baseline; fully correct runs number
\SkillsAllCorrectTreatmentPassed{}/\SkillsAllCorrectDenominator{} and
\SkillsAllCorrectBaselinePassed{}/\SkillsAllCorrectDenominator{}, respectively.
These results are descriptive of the evaluated model/provider configuration,
not population-level rates.}{}
The system does not claim to eliminate prompt injection, provide kernel-grade
sandboxing, roll back irreversible external effects, or make local evidence
cryptographically tamper-proof.  Its contribution is a runtime substrate for
agents whose affordances and state can change after deployment.
\end{abstract}

\maketitle

\section{Introduction}

LLM agents are moving from short chat loops toward long-running software
actors.  A coding or research agent may maintain task memory across sessions,
activate a reusable Skill, register a self-authored Deno/TypeScript JIT Tool,
fork a child process, execute a shell command, request Human approval,
checkpoint its state, resume after a failure, and later spawn or exec from a
checkpoint-derived image.  Research on self-evolving agents makes this trend
explicit: systems now adapt prompts, memory, tools, workflows, and agent code
both within and across task
episodes~\cite{gao2025selfevolving,wang2023voyager,zhang2024aflow,zheng2025skillweaver,qiu2025alita,zhang2025dgm}.

At the model--tool boundary, many agent systems implement variants of a
recurring loop: expose action schemas, obtain a model-selected action, dispatch
host functionality, return the result, and continue.  ReAct, Toolformer,
AutoGen, MetaGPT, SWE-agent, and related systems demonstrate the utility of
model-directed tool use and structured agentic
execution~\cite{yao2023react,schick2023toolformer,wu2023autogen,hong2023metagpt,yang2024sweagent}.
When treated as the primary runtime boundary, however, this interface can
conflate several questions that must remain distinct in a persistent,
self-modifying system:
\begin{enumerate}[leftmargin=1.4em]
  \item Which action descriptions are visible to the model?
  \item Which operation may this process invoke on which resource?
  \item Which labeled information may flow to which recipient?
  \item Which durable records explain what was attempted and what happened?
\end{enumerate}
Tool visibility answers only the first question.  A schema can guide a model's
action selection, but it is neither an unforgeable authority token nor a
data-flow policy, and it does not prove that an external effect committed.

This distinction matters in the presence of indirect prompt injection.  A
repository file or retrieved page can instruct an agent to activate a
``debugging'' Skill, register a generated Tool, send context to a remote
endpoint, and preserve the result in a reusable image.  Existing work shows
that ordinary-looking retrieved text can redirect tool-using agents toward
harmful actions~\cite{greshake2023not,zhan2024injecagent,ruan2024toolemu}.
Prompt defenses and planner--executor separation can reduce this attack
surface, but a runtime still needs a lower-level boundary that remains stable
as the model's action vocabulary changes.  Confirmation wrappers are
insufficient when the check resides in the wrapper rather than at the
resource-touching primitive.  Host isolation alone likewise neither attributes
an effect to the responsible process nor links the corresponding provider
outcome to the Capability, Task Authority ceiling, policy or Human approval,
and any exact release that admitted the operation.

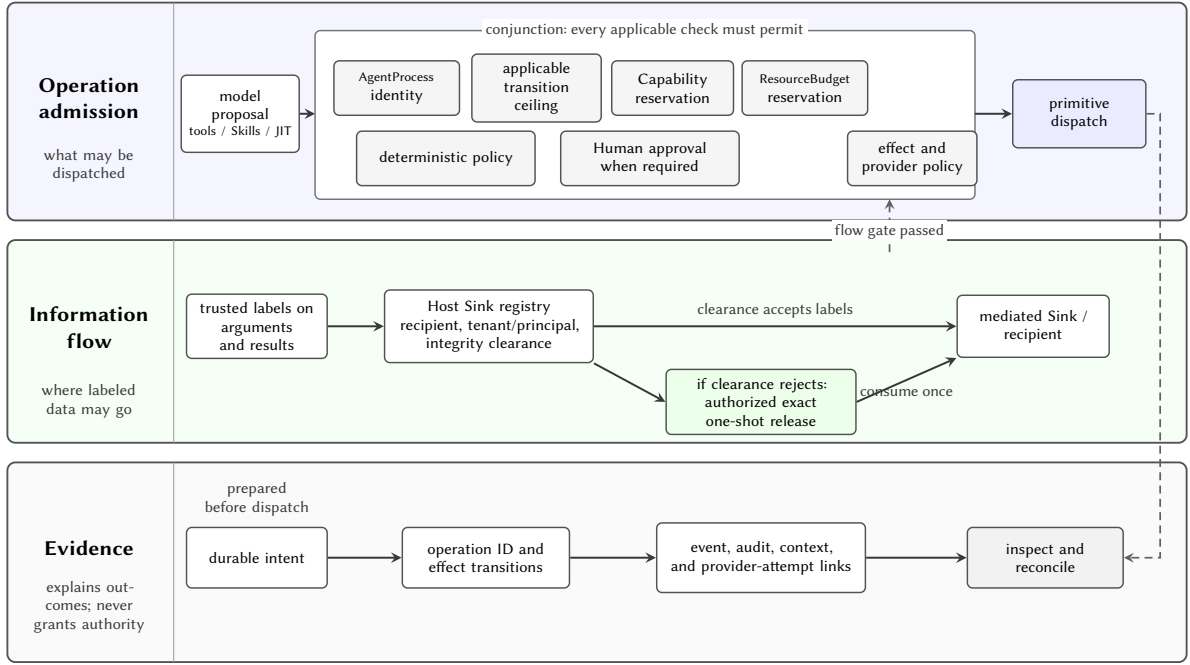
\begin{figure*}[t]
  \centering
\begin{tikzpicture}[
  x=1cm,
  y=1cm,
  >=stealth,
  font=\sffamily,
  box/.style={draw=black!68, fill=white, rounded corners=2pt,
    line width=0.55pt, align=center, inner sep=3pt,
    font=\sffamily\scriptsize},
  chip/.style={box, fill=black!4, font=\sffamily\scriptsize,
    minimum height=0.72cm},
  arrow/.style={->, line width=0.75pt, draw=black!78},
  cross/.style={->, densely dashed, line width=0.65pt, draw=black!62},
  note/.style={font=\sffamily\scriptsize, align=center, text=black!78}
]
  \path[use as bounding box] (0,0) rectangle (15.9,9.05);

  \draw[draw=black!68, fill=blue!4, rounded corners=3pt, line width=0.7pt]
    (0.15,6.02) rectangle (15.75,8.90);
  \draw[black!38] (2.35,6.02) -- (2.35,8.90);
  \node[align=center, font=\sffamily\small\bfseries, text width=1.85cm]
    at (1.23,7.64) {Operation\\admission};
  \node[note, text width=1.85cm] at (1.23,6.75)
    {what may be dispatched};

  \node[box, text width=1.35cm, minimum height=1.02cm] (proposal)
    at (3.23,7.43) {model proposal\\[-1pt]{\tiny tools / Skills / JIT}};
  \draw[draw=black!55, fill=white, rounded corners=2pt, line width=0.55pt]
    (4.22,6.30) rectangle (12.95,8.53);
  \node[note, fill=white, inner sep=1.5pt] at (8.58,8.53)
    {conjunction: every applicable check must permit};
  \node[chip, text width=1.45cm] (identity) at (5.30,7.77)
    {{\tiny AgentProcess}\\identity};
  \node[chip, text width=1.50cm] (taskauth) at (7.15,7.77)
    {applicable\\transition\\ceiling};
  \node[chip, text width=1.40cm] (capability) at (8.95,7.77)
    {Capability\\reservation};
  \node[chip, text width=1.45cm] (budget) at (10.70,7.77)
    {{\tiny ResourceBudget}\\reservation};
  \node[chip, text width=1.50cm] (effectpolicy) at (12.12,6.84)
    {effect and\\provider policy};
  \node[chip, text width=2.15cm] (detpolicy) at (5.95,6.84)
    {deterministic policy};
  \node[chip, text width=2.15cm] (human) at (8.65,6.84)
    {Human approval\\when required};
  \node[box, fill=blue!8, text width=1.55cm, minimum height=0.90cm] (dispatch)
    at (14.33,7.43) {primitive\\dispatch};

  \draw[arrow] (proposal.east) -- (4.22,7.43);
  \draw[arrow] (12.95,7.43) -- (dispatch.west);

  \draw[draw=black!68, fill=green!4, rounded corners=3pt, line width=0.7pt]
    (0.15,3.08) rectangle (15.75,5.76);
  \draw[black!38] (2.35,3.08) -- (2.35,5.76);
  \node[align=center, font=\sffamily\small\bfseries, text width=1.85cm]
    at (1.23,4.61) {Information\\flow};
  \node[note, text width=1.85cm] at (1.23,3.63)
    {where labeled data may go};

  \node[box, text width=1.65cm, minimum height=0.82cm] (labels)
    at (3.45,4.62) {trusted labels on\\arguments and results};
  \node[box, text width=2.55cm, minimum height=0.96cm] (sink)
    at (6.52,4.62) {Host Sink registry\\recipient, tenant/principal,\\integrity clearance};
  \node[box, fill=green!8, text width=2.30cm, minimum height=0.86cm] (release)
    at (10.12,3.62) {if clearance rejects:\\authorized exact one-shot release};
  \node[box, text width=1.80cm, minimum height=0.82cm] (recipient)
    at (13.72,4.62) {mediated Sink /\\recipient};

  \draw[arrow] (labels.east) -- (sink.west);
  \draw[arrow] (sink.east) -- node[note, above] {clearance accepts labels}
    (recipient.west);
  \draw[arrow] (sink.south east) -- (release.west);
  \draw[arrow] (release.east) -- node[note, below] {consume once}
    (recipient.south west);
  \draw[cross] (11.82,5.60) -- (11.82,6.30);
  \node[note, fill=white, inner sep=1pt] at (11.82,5.88) {flow gate passed};

  \draw[draw=black!68, fill=black!2, rounded corners=3pt, line width=0.7pt]
    (0.15,0.18) rectangle (15.75,2.82);
  \draw[black!38] (2.35,0.18) -- (2.35,2.82);
  \node[align=center, font=\sffamily\small\bfseries, text width=1.85cm]
    at (1.23,1.69) {Evidence};
  \node[note, text width=1.85cm] at (1.23,0.91)
    {explains outcomes; never grants authority};

  \node[box, text width=1.65cm, minimum height=0.80cm] (intent)
    at (3.45,1.56) {durable intent};
  \node[box, text width=2.00cm, minimum height=0.80cm] (transitions)
    at (6.48,1.56) {operation ID and\\effect transitions};
  \node[box, text width=2.55cm, minimum height=0.92cm] (links)
    at (10.12,1.56) {event, audit, context,\\and provider-attempt links};
  \node[box, fill=black!5, text width=1.85cm, minimum height=0.80cm] (inspect)
    at (13.88,1.56) {inspect and\\reconcile};

  \draw[arrow] (intent.east) -- (transitions.west);
  \draw[arrow] (transitions.east) -- (links.west);
  \draw[arrow] (links.east) -- (inspect.west);
  \draw[cross] (dispatch.east) -- (15.40,7.43) --
    (15.40,1.56) -- (inspect.east);
  \node[note, text width=1.55cm] at (3.45,2.34)
    {prepared before dispatch};
\end{tikzpicture}
  \caption{The three-plane runtime boundary.  Mutable model-visible
  affordances determine which operations the model can request; operation
  admission and information-flow admission remain distinct gates at concrete
  primitives, each enforced as a conjunction of Host-side checks.  Durable
  evidence records and links decisions without conferring authority.}
  \Description{Three horizontal planes. The operation plane conjoins process,
  task, capability, policy, budget, and primitive checks; the information-flow
  plane compares labels with Host-established Sink clearance or an exact
  one-shot release; and the evidence plane records intent and outcomes without
  granting authority.}
  \label{fig:three-plane}
\end{figure*}

\sys starts from a narrow systems principle:
\begin{quote}
\emph{A self-evolving agent may change what it can ask for, but it cannot
thereby change what it is authorized to affect or where its information may
flow.}
\end{quote}
The design expresses this principle through two admission planes and a separate
evidence plane (Figure~\ref{fig:three-plane}).  Operation admission requires the
conjunction of process identity, the applicable Task Authority ceiling,
Capabilities, deterministic policy or Human approval, resource accounting, and
a typed primitive.  Information-flow admission propagates Runtime-maintained
labels and immutable source references, resolves a canonical Sink through the
Host-owned registry, checks its clearance, and, for conditionally permitted
high-sensitivity egress, requires an exact one-shot Human release.  The evidence
plane links prepared intent, dispatch, classified outcomes, resource accounting,
events, and audit records, but cannot make a denied action admissible.

This paper makes four contributions.
\begin{enumerate}[leftmargin=1.4em]
  \item \textbf{A three-plane runtime model.}  We distinguish mutable
  affordances from resource authority, information-flow permission, and
  evidence.  We state admission and recovery invariants for persistent,
  self-evolving processes.
  \item \textbf{Conjunctive admission and exact release.}  We combine typed
  Capability checks with applicable Task Authority ceilings, policy or Human
  approval, hierarchical budgets, trusted data labels, a Host-owned Sink
  registry, and exact one-shot release.  Neither plane can supply a missing
  predicate in the other.
  \item \textbf{Crash-aware protected effects.}  Provider-backed operations
  durably prepare intent and reserve finite authority before dispatch, classify
  outcomes as terminal or ambiguous, and, when a provider contract supplies an
  explicit reconciliation hook, reconcile receipts upon reopening without
  blind replay.  The Explainable Operations layer links this protocol to LLM,
  Tool, syscall, primitive, Capability, Human, provider-effect, resource, event,
  and audit evidence.
  \item \textbf{A source-bound evaluation design.}  The artifact separates
  deterministic authority tests, native-live and modeled practical workflows,
  structural recovery gates, a canonical real-model durable-workflow gate, and
  a paired comparison of Skill projection with a full-schema baseline into
  distinct evidence classes.  Every quantitative claim is generated from
  hashed reports bound to one clean source revision; real-model reports are
  additionally bound to their provider configuration.
\end{enumerate}

The claim is architectural rather than planner-centric.  \sys proposes neither
a new reasoning prompt nor an optimizer, and it does not assume that the model
reliably recognizes malicious instructions.  It aims to mediate resource
effects and runtime-delivered information flows through explicit Host-composed
boundaries.

\section{Threat Model and System Model}
\label{sec:model}

\subsection{Assets, adversaries, and trusted components}

Protected assets include filesystem and Git state; shell and PTY execution;
Object Memory and prompt context; Human interactions; process and image
lifecycles; checkpoints; provider credentials and configuration; registered
remote providers and the systems they reach; LLM calls; finite resource
budgets; labeled data; and the durable records required to explain or recover
operations.  The adversary may control model output, repository files,
retrieved documents, Tool results, generated Skill text, JIT source,
remote-provider responses, Human free-text answers, process messages, or other
content presented to a model.  It may know Tool names, Object identifiers,
endpoint identifiers, namespace strings, checkpoint identifiers, and resource
paths.  It may induce the model either to request a dangerous action or to
compose several actions that appear individually plausible but are dangerous
in combination.

The trusted computing base (TCB) comprises Runtime managers, Host-side
composition logic, trusted issuers, the Human operator when an approval or
release is accepted, the configured Resource Provider Substrate, accepted
Runtime Modules and their
source-identity bindings, registered provider policies, the Python interpreter,
the storage backend, and the host OS\@.  Each provider is trusted only to the
extent required by its declared boundary.  For example, an MCP stdio executable
runs as native code under the Host account.  Skills, JIT code, model output,
workspace-supplied image definitions, repository contents, and remote payloads
are not trusted to enforce policy.  An administrator with direct database
access can alter local evidence and therefore remains within the TCB\@.  When the
bundled GUI is enabled, its Electron main/preload/renderer stack, packaged
assets, and dependency supply chain also enter the TCB because they can reach
Host/operator surfaces.

The system is designed to mediate operations and information delivered through
the Runtime.  It does not claim to prevent untrusted text from changing a
model's plan; prevent a trusted recipient from forwarding data after delivery;
encrypt data at rest; provide a kernel, hypervisor, or verified sandbox;
compensate for irreversible provider effects; or expose a public remote
administration API\@.  These scope boundaries matter: a Capability denial is
evidence only that a particular mediated action was denied, not that every
possible effect on the host is impossible.

\subsection{Runtime state}

We model each process as a runtime subject with state
\begin{equation}
\begin{split}
p = \langle & \mathit{id}, \mathit{img}, \mathit{status}, \mathit{mem},
               \mathit{cwd}, T, S, J,\mathcal{C},\mathcal{M},\\
             & \mathcal{B},\mathcal{L},Q,\mathit{children} \rangle.
\end{split}
\label{eq:process-state}
\end{equation}
Here, $\mathit{id}$ identifies the process; $\mathit{img}$ is its \agentimage;
and $\mathit{status}$ is its lifecycle state.  The variables $\mathit{mem}$ and
$\mathit{cwd}$ denote process-local views of
Object Memory and the working directory, respectively.  The variables $T$,
$S$, and $J$ denote the callable Tool table, loaded Skills, and registered JIT
Tools; $\mathcal{C}$ is the Capability set; $\mathcal{M}$ is the active Task
Authority Manifest; $\mathcal{B}$ is the hierarchical budget state;
$\mathcal{L}$ contains the labels and source references associated with the
process's current data; $Q$ is a durable message queue; and
$\mathit{children}$ records the process-tree relation.  Explicit lifecycle
states represent spawn, fork,
exec, wait, signal, pause, resume, and exit operations instead of encoding them
as chat metadata.

An \agentimage defines the default boot-prompt composition, the complete
initial Tool table, the narrower initial model projection, default Skills and
JIT candidates, LLM-profile selection, context policy, required startup
modules, and optional workspace materialization.  Required Capabilities
declared by an ordinary image serve as diagnostics for Host review, not as
grants or launch gates.  The narrow image-package bootstrap exception may
materialize a fresh private workspace and issue only the filesystem grants
specified by the Host-accepted package manifest, and only over that copy.  It
grants no authority over an existing workspace.  Restoring a checkpoint cannot
mint authority declared by the image.

The \emph{model-visible action surface} of $p$ is
\begin{equation}
A_p = V(T_p,S_p,J_p,I_p,K_p,R_p),
\label{eq:action-surface}
\end{equation}
where $V$ computes the current model projection; $I_p$ and $K_p$ denote image
and checkpoint affordances; and $R_p$ denotes registered remote affordances.
The complete callable table is maintained separately from $V$: a Skill can
project a bounded subset of built-in Tools on demand, while entries hidden from
the model remain callable by trusted direct workflows.

Resource authority is distinct from $A_p$.  It comprises the operations
admitted by $\mathcal{C}_p$ under typed resource matching, effects, constraints,
expiration, finite-use limits, and delegation lineage.  For an explicit Host
manifest and any manifest derived from it, a durable
\code{transition_ceiling} further intersects that authority during
agent-controlled transitions.  Let $\mathcal{C}^{\mathrm{ext}}_p$ denote
authority over pre-existing or external resources.  Writing $\Delta^+$ for an
expansion of a set or a relaxation of a ceiling,
\begin{equation}
\Delta^+ A_p \not\Rightarrow \Delta^+\mathcal{C}^{\mathrm{ext}}_p
\quad\land\quad
\Delta^+ A_p \not\Rightarrow \Delta^+\mathcal{M}_p.
\label{eq:no-implicit-authority}
\end{equation}
Skill activation, JIT registration, image construction, child creation,
checkpoint restoration, and endpoint discovery can change $A_p$, but none of
these transitions implicitly expands authority over pre-existing or external
resources.  When a process creates a new Object, the Runtime may issue the
creator a handle to that newly created, owned resource.  The trusted
image-package bootstrap may similarly issue the bounded private-workspace
grants described above.  These narrowly scoped fresh-resource grants do not
constitute ambient permission administration.

\subsection{Three planes}

For a normalized operation request $r$ issued by $p$, the intended
operation-admission predicate is
\begin{equation}
\begin{split}
\mathsf{OpAdmit}(p,r) \equiv {}&
\mathsf{live}(p) \land \mathsf{cap}(\mathcal{C}_p,r) \\
&\land\ \mathsf{ceiling}(\mathcal{M}_p,r)
\land \mathsf{policyOrHuman}(p,r) \\
&\land\ \mathsf{budget}(p,r) \land \mathsf{primitive}(r).
\end{split}
\label{eq:operation-admission}
\end{equation}
All predicates must hold.  An \code{allow} Capability cannot exceed an
applicable Task Authority ceiling; Human approval cannot compensate for a
missing Capability; a budget does not authorize the operation it measures; and
resource identity is normalized and checked at the primitive boundary.  The
$\mathsf{ceiling}$ predicate imposes a restriction when $\mathcal{M}_p$ is an
explicit \code{transition_ceiling} manifest or is derived from one.  Every
launch also receives a durable implicit compatibility record.  That record
binds launch and model-request evidence but does not constrain authority that
the Host deliberately grants later.

For runtime-mediated delivery of payload descriptor $x$ to canonical Sink $s$,
let $\mathcal{V}$ denote the Host-resolved source-version set and let $\ell$
denote its aggregate labels.  Information-flow admission is
\begin{equation}
\begin{split}
\mathsf{FlowAdmit}(p,r,x,s) \equiv {}&
\mathsf{sources}(x)=\mathcal{V} \land
\mathsf{labels}(\mathcal{V})=\ell \\
&\land\left(\mathsf{clear}(s,\ell)\ \lor{}\right.\\[-0.35em]
&\qquad\left.\mathsf{consume}
  (\mathsf{release}(p,x,s,\mathcal{V},\ell))\right)\\
&\land\ \mathsf{integrity}(\ell) \succeq
\mathsf{minIntegrity}(r).
\end{split}
\label{eq:flow-admission}
\end{equation}
The Host, not the model, establishes Sink identity and trust.  Labels record
sensitivity, tenant and principal identities, trust, and integrity.  An exact
release is bound to the actor, Sink, payload identity, source versions, label
set, Task Authority Manifest, and Sink-registry generation.  The release is
consumed by the corresponding dispatch and leaves the source labels unchanged.
The minimum integrity requirement is independent of sensitivity clearance:
neither trusting a Sink nor granting a sensitivity release can endorse
low-integrity data.

Internal process handoff is deliberately not treated as delivery to a cleared
external Sink.  Messages retain their labels and must satisfy the receiving
process's Task Authority receive domain.  A child or sibling therefore cannot
serve as an implicit declassifier.

The evidence plane comprises causal operations $O$, events $E$, audit records
$D$, provider-effect state $F$, context manifests $X$, and resource settlement
$U$.  None of these records is an admission credential:
\begin{equation}
\forall z \in O\cup E\cup D\cup F\cup X\cup U,\quad
\neg\mathsf{credential}(z).
\label{eq:evidence-not-authority}
\end{equation}
Evidence may support or refute a claim, diagnose a failure, or aid
reconciliation; its presence may also show that admission occurred earlier.
It cannot itself satisfy a missing admission predicate.

\subsection{Safety invariants}

We structure the implementation around four invariants that can be checked at
primitive and recovery boundaries.

\textbf{I1: no implicit authority over existing resources.}
Equation~\ref{eq:no-implicit-authority} holds for every agent-controlled
transition that changes affordances.  Child delegation attenuates authority.  A
checkpoint fork intersects restored authority with both current source
authority and any applicable explicit or derived Task Authority ceiling;
revocations override stale restored state.  Authority may nevertheless grow
through deliberate later Host grants under an implicit root record, handles for
newly created Objects, or the trusted private-workspace bootstrap.  In each
case, the growth results from an explicit Host or Runtime policy decision, not
from growth of the model-visible action surface.

\textbf{I2: conjunctive dispatch.}  The design requires every provider dispatch
to satisfy Equation~\ref{eq:operation-admission}.  If the operation's contract
declares egress, the dispatch must also satisfy
Equation~\ref{eq:flow-admission}.  The Runtime
revalidates all state-dependent bindings at dispatch time rather than relying
solely on an earlier preview.  Section~\ref{sec:evaluation} reports the bound
artifact's coverage and known exception to this requirement.

\textbf{I3: durable intent and terminal linkage.}  Each logical protected
operation establishes exactly one durable intent identity before its first
provider dispatch; all ordered provider phases share that identity.  Every
terminal outcome finalized by the Runtime includes all contract-required
transition, event, audit, causal, authority/release, and resource links.  A
crash before finalization leaves a pending or unknown recovery input rather
than a fabricated terminal record.  Evidence completeness is a provenance
property: it is neither a safety score nor proof that the provider behaved
honestly.  We exclude from \emph{provider dispatch} the single checker-pinned
Semantic Git preflight exception.  Before ordinary operation authority has
been established and an intent can be created, this trusted local preflight may
perform only a read-only security observation.  It returns bounded digests and
flow metadata; a subsequently admitted Git operation repeats the observation
within the ordinary Protected Operation.

\textbf{I4: reopen is not redispatch.}  If a crash leaves an effect dispatched
but without a certified terminal outcome, its status remains unknown.  Upon
reopening, the Runtime may repair locally prepared state.  It may query a
stable provider receipt only if the corresponding contract installs an
explicit reconciliation hook.  It may neither infer non-execution from the
absence of a return value nor blindly replay the provider call.  A linked Task
Run may independently project the unresolved effect as
\code{needs_attention}.

\section{Authority and Information-Flow Admission}
\label{sec:admission}

\subsection{Capabilities and Task Authority}

A \capability identifies a typed resource, a set of rights, an effect
($\mathsf{allow}$, $\mathsf{ask}$, or $\mathsf{deny}$), constraints, issuer and
holder identities, optional expiration and a finite-use count, and delegation
lineage.  Resources are typed identifiers rather than free-form wrapper labels.
Examples include workspace-relative filesystem paths, pinned repository
identities and Git operations, Object Memory namespaces and Objects, process
and image subjects, registered JSON-RPC methods and MCP items, shell policies,
Human channels, and checkpoints.  Filesystem and Git identities are
canonicalized before authorization.  JSON-RPC and MCP use a two-stage,
metadata-hiding gate: public endpoint or server identifiers and public item
identifiers first undergo a coarse invocation-authority check.  Only an
authorized caller may resolve hidden registry metadata; the primitive then
performs exact authorization and revalidates the captured specification digest
and registry generation immediately before dispatch.  Explicit deny overrides
allow, and delegation may preserve or narrow, but never widen, a parent grant.

Capabilities determine whether a subject may request an operation on a
resource; they do not determine whether that operation falls within the current
task.  An explicit Host-authored \taskauthority establishes a durable
\code{transition_ceiling}; every manifest derived from it preserves or narrows
that ceiling for the corresponding process.  It bounds launch authority and
constrains permitted provider effects, data receive domains, resource-budget
ceilings, approval policy, and any configured semantic-auto-approval rules for
the process and subsequent agent-controlled child, fork, exec, and
permission-request transitions.  The manifest is canonicalized, and its
unkeyed content hash is bound into approval and operation evidence.  If the
Host omits an explicit template, the Runtime instead persists an implicit
compatibility manifest: Capabilities granted at launch still bound subsequent
model permission requests, but a deliberate later Host grant may extend beyond
that record.  An image's \code{required_capabilities} field is a declaration
and diagnostic only.  The Runtime records unsatisfied requirements, but they
neither block launch nor create live authority.

Where a transition ceiling applies, the two mechanisms are intersected.  A
broad reusable Capability cannot exceed a narrow task-effect ceiling, while a
permissive manifest cannot create resource authority.  For schema-v2 effect
ceilings, an omitted field or JSON \code{null} imposes no additional
restriction, whereas \code{[]} denies all provider effects.  Legacy schema-v1
rows are upcast while preserving their historical interpretation of an empty
list as unrestricted.  Canonicalization preserves these distinctions
explicitly rather than relying on truth-value coercion.

Human approval supplies a further conjunct; it is not an ambient source of
authority.  An \code{ask} decision creates a typed pending request with a
canonical preview.  A terminal response explicitly selects
\code{always_allow}, \code{always_deny}, or \code{ask_each_time}, subject to
Host constraints; exact one-time authority instead uses a separate one-use
Capability lease.  On resume, the Runtime re-enters the original primitive and
revalidates its actor, resource, canonical arguments, manifest, and target-state
bindings.  Neither a textual response nor a GUI action is converted into a
general-purpose Capability.

\subsection{Budget admission and settlement}

Hierarchical budgets account for logical Tool and LLM calls, filesystem
read/write bytes, JSON-RPC and MCP request/response bytes, Deno syscalls,
subprocess wall time, CPU time, peak RSS, and other Runtime resources.
Each charge is applied along the full acting-process-to-ancestor accounting
chain.  Budget admission limits how much otherwise-admissible work may run; it
does not authorize that work.

The Runtime distinguishes a nonreserving preflight check from a durable
reservation.  Preflight rejects an operation that cannot fit but holds no
quota.  A reservation instead records a maximum usage envelope before dispatch
and later settles observed usage against that envelope.  A
certified-not-started outcome releases the reservation; an ambiguous
post-dispatch outcome is maximum-settled when no better measurement is
available.  Effect and evidence finalization may precede resource settlement.
Consequently, an accounting failure after an external effect cannot undo that
effect; the unresolved reservation remains durable input to recovery.  This
ordering avoids claiming transactional atomicity across a non-transactional
provider.

Each executor-level logical LLM call reserves one call together with the
configured maximum envelope for prompt, completion, and total tokens before
provider dispatch.  Bounded transport attempts, retry/continuation state, and
provider-reported usage are recorded as observations within that logical call.
These counters therefore do not represent exact physical-request counts,
provider billing, or monetary spend.

\subsection{Labels, source identity, and Sinks}

Runtime-mediated payloads carry a \emph{DataFlowContext} that associates labels
with immutable source references.  Aggregating contexts selects the highest
sensitivity, the lowest integrity and trust, and a conservative join of
tenant/principal identities.  Versioned Objects and file bindings contribute
immutable references, while Human responses, LLM/provider ingress, Tool
results, JIT Tool results, process messages, JSON-RPC and MCP responses, and typed Git
observations propagate Runtime-owned flow context.  Joins remain conservative:
the Runtime never drops a label merely because the model appears not to have
used some part of its context.

Each external egress targets a concrete Sink.  The Host owns a durable,
canonical Sink registry whose generation advances on every mutation.  Under
the default configuration, unmatched Sinks are untrusted with a \code{normal}
sensitivity ceiling; more generally, no untrusted Sink can clear data above
\code{normal}.  A Sink rule specifies a trust level, maximum sensitivity,
tenant/principal allowlists, and, for provider-backed clearance above
\code{normal}, a provider identity hash.  An operation contract may
additionally impose an independent minimum source-integrity floor.  The model
cannot assert in a Tool argument that a URL, shell, PTY, filesystem target,
provider, or GUI surface is trusted.  An operation with multiple actual
recipients must declare every corresponding Sink; authorizing only a convenient
primary recipient is insufficient.

Information-flow enforcement spans Runtime-mediated boundaries for LLM, Human,
JSON-RPC, MCP, typed Git, filesystem writes, shell/PTY input and control,
process handoff, and selected GUI presentation.  Because a filesystem read or clock
observation may receive information without transmitting the caller's payload,
the operation contract explicitly declares its direction
($\mathsf{none}$, $\mathsf{ingress}$, $\mathsf{egress}$, or
$\mathsf{bidirectional}$); the Runtime does not infer direction from the legacy
\code{information_flow} flag.

\subsection{Exact release and conservative declassification}

When a \code{conditional} Sink requires sensitivity clearance above
\code{normal} and the requested sensitivity remains within the rule's maximum,
the Runtime may create a typed Human request for an exact one-shot release.
The release binds the process and operation; Sink and trust identities and the
registry generation; the canonical payload or argument digest; source Object
and file versions; effective labels; the Task Authority Manifest hash; any
target state; and a deadline.  It is nondelegable and single-use.
The preparation phase of a Protected Operation reserves the release, and
successful settlement consumes it.  A certified-not-started path may restore the same live
reservation, whereas an unknown provider outcome consumes the authority and
prohibits automatic replay.

An exact release does not mutate labels: the source remains labeled, later
payloads undergo independent checks, and minimum-integrity floors remain
nonwaivable.  Lowering an Object's sensitivity, removing or replacing its
identity, raising its trust or integrity, or changing its declassification
authority requires an exact \code{declassification:object:<oid>} Capability
with the \code{admin} right.  Bundled model-facing memory Tools cannot request
this transition; a process may perform it only after the Host deliberately
issues the Capability.  This high-water design differs from permissive IFC
analyses that identify influential inputs and regenerate outputs under reduced
contexts to derive less restrictive labels~\cite{siddiqui2025permissive}, and
from planner-level architectures based on trusted control flow or deterministic
label propagation~\cite{debenedetti2026camel,costa2025fides}.  By design, \sys
preserves high-water labels and confines exceptional egress to narrow,
inspectable releases; it does not claim a general noninterference theorem.

\subsection{Semantic Shadow is not an allow oracle}

An optional semantic plane can assess eligible approval requests and committed
provider ingress, record payload-free FlowGraph evidence, and emit conservative
findings.  It is disabled by default.  In shadow mode, it records
counterfactual outcomes but cannot change decisions.  In enforcement modes,
machine denial is limited to a closed set of deterministic Host violations.
Only the canary path may issue an exact, short-lived, nondelegable, one-use
Capability for a frozen catalog of low-risk read and diff operations, and only
when every independent Host predicate is already satisfied.

The classifier returns only closed, typed findings together with confidence,
calibration, out-of-distribution, and abstention metadata; its schema has no
general \code{allow} or \code{deny} field.  Findings may veto candidate canary
issuance or require Human review, but cannot themselves cause a terminal denial,
select a rule, mutate labels, widen a ceiling, or supply a missing allow
predicate.  The Host-only settlement path re-reads the pending request and
revision, live Task Authority, DataFlow and FlowGraph state, policy epoch,
binding and target state, classifier provenance, and budget within a Host
transaction.  Learned semantic analysis thus provides only fallible evidence;
admission remains deterministic and Host-controlled.

\section{Protected Operations and Causal Evidence}
\label{sec:protected}

\subsection{Why provider calls need a protocol}

Filesystem, Git, shell, LLM, JSON-RPC, MCP, Human, and PTY providers can produce
effects outside the enclosing RuntimeStore transaction.  A process may crash
after the provider commits an effect but before the caller receives a return
value; conversely, it may crash after local preparation but before dispatch.
Treating both cases as ordinary Tool-call failures can either discard evidence
or cause a retry to duplicate an external effect.

The \protectedop SDK is the stable extension boundary for provider-backed host
operations.  A trusted contract declares the authority mode, provider and
operation identities, effect and resource policies, information-flow
direction, minimum integrity, required evidence roles, and any prepared-state
recovery handler.  The domain primitive remains responsible for argument
normalization and exact Capability decisions, while the SDK governs the common
lifecycle and its fail-closed evidence contract.

\subsection{Prepare, dispatch, and settlement}

For an invocation $r$, the protocol proceeds as follows:
\begin{enumerate}[leftmargin=1.4em]
  \item \textbf{Normalize and preflight.}  The primitive resolves the concrete
  resource, canonicalizes the arguments, and identifies the actor, provider
  target, any applicable Task Authority ceiling, Sinks, source references, and
  optional resource envelope.
  \item \textbf{Prepare atomically.}  In a single RuntimeStore transaction, the
  SDK reauthorizes and reserves both finite Capability uses and exact-release
  uses, creates a durable usage reservation when the invocation supplies a
  resource envelope, creates the external-effect intent, and links that intent
  to the causal operation.  No provider code has run at this point.
  \item \textbf{Revalidate and dispatch.}  Immediately before each provider
  phase, the SDK rechecks all reusable or reserved authority and every declared
  mutable binding: the semantic control epoch, provider-registry identity, Sink
  identity, target-state resolver, and DataFlow sources.  A contract such as
  typed Git may also bind a target-specific state token.  The SDK then crosses
  the provider boundary using only Host-configured transport and credentials.
  \item \textbf{Classify and finalize.}  Successful and failed completions are
  classified and finalized as $\mathsf{committed}$, $\mathsf{failed}$, or
  $\mathsf{unknown}$, with a rollback class, status, and bounded observations.
  Certified-not-started outcomes are handled separately: when
  no earlier phase has crossed the effect/authority floor, the SDK restores the
  exact reservations and abandons the prepared intent.  Normally, the terminal
  effect transition, event, audit record, and causal links are committed before
  the SDK returns.  The narrow \code{PRESERVE_RESULT} exception for Human
  read/write operations may return a provider-accepted answer or delivery even
  if local completion fails, but only after conservative fallback settlement
  succeeds.  The surviving effect remains pending or unknown, and the Human
  provider is not invoked again.
  \item \textbf{Settle resources.}  The reservation is settled against observed
  usage.  A certified-not-started outcome releases finite authority, whereas
  dispatched or unknown paths consume it.  A settlement failure cannot undo a
  committed provider effect.
\end{enumerate}

\begin{figure*}[t]
  \centering
\begin{tikzpicture}[
  x=1cm,
  y=1cm,
  >=stealth,
  font=\sffamily,
  box/.style={draw=black!68, fill=white, rounded corners=2pt,
    line width=0.55pt, align=center, inner sep=3pt,
    font=\sffamily\scriptsize},
  state/.style={box, minimum height=0.60cm, font=\sffamily\scriptsize},
  arrow/.style={->, line width=0.75pt, draw=black!78},
  recovery/.style={->, densely dashed, line width=0.75pt, draw=black!68},
  note/.style={font=\sffamily\scriptsize, align=center, text=black!78}
]
  \path[use as bounding box] (0,0) rectangle (15.9,9.55);

  \begin{scope}[yshift=0.75cm]

  \node[box, fill=blue!4, text width=1.15cm, minimum height=0.78cm] (request)
    at (0.82,6.86) {effect\\request};

  \draw[draw=black!65, fill=blue!3, rounded corners=3pt, densely dashed,
    line width=0.7pt] (1.72,5.45) rectangle (7.15,8.32);
  \node[note, font=\sffamily\scriptsize\bfseries, fill=blue!3, inner sep=1.5pt]
    at (4.43,8.32) {RuntimeStore transaction A (atomic)};
  \node[box, text width=1.52cm, minimum height=0.80cm] (validate)
    at (2.72,7.10) {validate\\authority, flow,\\and budget};
  \node[box, text width=1.58cm, minimum height=0.80cm] (reserve)
    at (4.70,7.10) {reserve finite\\Capability, release,\\and resources};
  \node[box, fill=blue!8, text width=3.48cm, minimum height=0.74cm] (prepare)
    at (4.43,5.98) {write durable \texttt{prepared} intent,\\operation ID, and idempotency key};
  \draw[arrow] (validate.east) -- (reserve.west);
  \draw[arrow] (reserve.south) -- (prepare.north east);
  \draw[arrow] (request.east) -- (1.72,6.86);

  \node[box, fill=yellow!9, text width=1.72cm, minimum height=0.88cm] (dispatch)
    at (8.48,6.86) {dispatch to\\configured provider};
  \draw[arrow] (7.15,6.86) -- node[note, above] {commit A} (dispatch.west);
  \node[note, font=\sffamily\scriptsize\bfseries, text width=2.2cm]
    at (8.48,8.02) {outside the storage transaction};
  \node[note, text width=1.70cm] at (7.38,5.60)
    {C1: crash after prepare};
  \node[note, text width=1.70cm] at (9.30,5.17)
    {C2: crash/timeout around dispatch};

  \node[state, fill=green!8, text width=1.25cm] (committed)
    at (10.87,7.77) {verified\\committed};
  \node[state, fill=red!5, text width=1.25cm] (failed)
    at (10.87,6.86) {verified\\failed};
  \node[state, fill=yellow!10, text width=1.25cm, line width=0.85pt] (unknown)
    at (10.87,5.62) {unknown\\outcome};
  \draw[arrow] (dispatch.east) -- (committed.west);
  \draw[arrow] (dispatch.east) -- (failed.west);
  \draw[arrow] (dispatch.east) -- (unknown.west);

  \draw[draw=black!65, fill=green!3, rounded corners=3pt, densely dashed,
    line width=0.7pt] (12.05,5.88) rectangle (15.72,8.27);
  \node[note, font=\sffamily\scriptsize\bfseries, fill=green!3, inner sep=1.5pt]
    at (13.88,8.27) {RuntimeStore transaction B};
  \node[box, fill=white, text width=2.90cm, minimum height=0.88cm] (settle)
    at (13.88,7.18) {append committed/failed transition;\\link outcome, event, and audit evidence};
  \node[note, text width=2.85cm] at (13.88,6.23)
    {resource settlement may follow effect commit};
  \draw[arrow] (committed.east) -- (12.05,7.77);
  \draw[arrow] (failed.east) -- (12.05,6.86);
  \end{scope}

  \node[box, text width=1.66cm, minimum height=0.72cm] (reopen)
    at (10.87,5.18) {reopen /\\recovery pass};
  \node[box, fill=yellow!8, text width=3.35cm, minimum height=0.86cm,
    line width=0.85pt] (hookcheck) at (10.87,4.00)
    {explicit provider hook\\\texttt{reconcile\_external\_effect}\\is configured?};
  \draw[recovery] (unknown.south) -- (reopen.north);
  \draw[recovery] (reopen.south) -- (hookcheck.north);

  \node[box, fill=green!4, text width=3.55cm, minimum height=0.88cm] (hookresult)
    at (7.18,2.62) {invoke hook; an authoritative\\terminal receipt may settle\\via transaction B};
  \node[box, fill=yellow!10, text width=2.75cm, minimum height=0.88cm,
    line width=0.85pt] (staysunknown) at (13.60,2.62)
    {external effect remains\\\texttt{unknown}; retain relevant\\reservations};
  \draw[recovery] (hookcheck.south west) -- node[note, above left] {yes}
    (hookresult.north east);
  \draw[recovery] (hookcheck.south east) -- node[note, above right] {no}
    (staysunknown.north west);
  \draw[recovery] (hookresult.south) -- (7.18,1.94) --
    node[note, below, text width=2.55cm] {no authoritative terminal result}
    (12.05,1.94) -- (staysunknown.south west);

  \node[box, fill=red!4, text width=2.90cm, minimum height=0.74cm] (taskprojection)
    at (13.60,1.24) {linked TaskRun projection only:\\\texttt{needs\_attention}};
  \draw[recovery] (staysunknown.south) -- node[note, right] {project blocker}
    (taskprojection.north);

  \draw[draw=black!78, fill=black!3, rounded corners=2pt, line width=0.85pt]
    (0.30,1.22) rectangle (4.70,2.16);
  \node[align=center, font=\sffamily\scriptsize\bfseries, text width=4.05cm]
    at (2.50,1.69) {No blind replay: reopen never redispatches an\\ambiguous external effect.};

  \draw[draw=black!62, fill=blue!3, rounded corners=2pt, line width=0.60pt]
    (0.30,0.18) rectangle (10.98,1.02);
  \node[align=center, font=\sffamily\tiny, text width=10.10cm]
    at (5.64,0.60) {Narrow Semantic Git exception: a checker-pinned Host call graph may make local read-only pre-intent observations only;\\it cannot mutate or contact a remote, creates no effect intent, and is not evidence of protected dispatch.};
\end{tikzpicture}
  \caption{Protected Operation lifecycle and crash windows.  Preparation binds
  authority, flow, resource, and evidence before provider dispatch.  An unknown
  outcome means that the provider may have acted.  A contract may reconcile
  this state through an explicitly installed hook, but reopening never grants
  replay permission.}
  \Description{A state machine with an atomic prepare transaction, provider
  dispatch outside that transaction, committed and failed terminal settlements,
  and an unknown branch. An explicit contract hook may reconcile the unknown
  branch; otherwise, it remains unknown. A linked Task Run may project that
  outcome as requiring attention.}
  \label{fig:protected-operation}
\end{figure*}
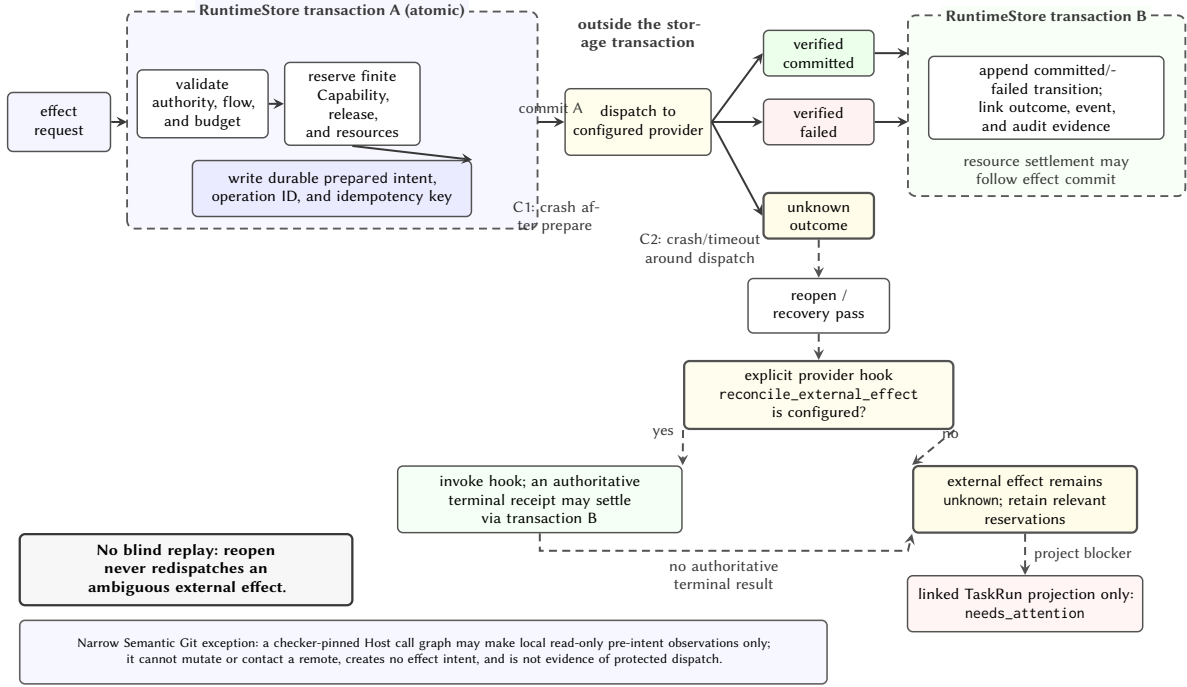

Figure~\ref{fig:protected-operation} shows the state machine.  A prepared-state
recovery callback may repair only local data within the recovery transaction;
it cannot call a provider.  Any contract that mutates local state during
preparation must name a handler installed before the Runtime is opened;
otherwise, startup fails closed.

The dispatch property is intentionally weaker than an end-to-end commit
property:
\begin{equation}
\begin{split}
\mathsf{providerDispatch}(f) \Rightarrow{}& \exists p,r,o.\\[-0.2em]
&\mathsf{prepared}(f,p,r,o) \land \mathsf{OpAdmit}(p,r) \\
&\land\bigl(\neg\mathsf{egress}(r)\lor{}\bigr.\\[-0.35em]
&\qquad\bigl.\mathsf{FlowAdmit}(p,r,x_r,s_r)\bigr) \\
&\land\mathsf{linked}(f,o,\mathsf{intent}).
\end{split}
\label{eq:protected-dispatch}
\end{equation}
Here, $x_r$ and $s_r$ denote the request's payload descriptor and canonical Sink.
Only Runtime finalization entails terminal evidence:
\begin{equation}
\begin{split}
\mathsf{finalized}(f,\mathsf{committed}) \Rightarrow{}& \exists o.\,\bigl(
\mathsf{classified}(f,\mathsf{committed})\\
&\land\mathsf{linked}(f,o,\mathsf{event},\mathsf{audit})\bigr).
\end{split}
\label{eq:protected-finalize}
\end{equation}
Thus, a crash after provider dispatch but before finalization may leave a
durable intent without its terminal event and audit links.  On reopening, the
Runtime must classify the outcome as unknown unless an explicitly installed
reconciliation hook supplies sufficient provider evidence.  This property is a
mediation invariant, not an exactly-once theorem.  A malicious or misconfigured
provider may violate its contract, and a direct Host bypass falls outside the
threat model.  For cooperating providers, stable idempotency keys and receipts
improve reconciliation without altering the conservative treatment of missing
evidence.

\subsection{Crash ambiguity and recovery}

The Runtime distinguishes three crash-relevant cases.  First, if no durable
intent exists, no effectful provider phase was entered through the operation.
The checker-pinned Semantic Git security observation described above is a
local, read-only preflight and therefore does not count as such a phase.  Second, if
an intent remains prepared and provider dispatch is certified not to have
started, recovery can repair local preparation state, abandon the intent, and
release reservations.  Third, if dispatch may have begun, recovery preserves
or consumes finite authority, charges the conservative resource envelope when
necessary, and leaves the outcome unknown.  Only a contract with an explicit
reconciliation hook may use a stable provider receipt or another bounded
observation to prove a terminal state.  A generic effect does not have a
\code{needs_attention} state; a linked Task Run may nevertheless project an
unresolved unknown effect as that operator-visible status.

Provider reconciliation, when explicitly supported, may restore an effect-bound
Capability reservation only in the narrowly defined certified-not-started
case.  This dependency explains why stale Capability reservations cannot be
abandoned before provider effects are reconciled at startup.  A missing
response, a timeout, or process death never constitutes evidence of
non-execution.

Checkpoint restore and image commit do not roll back external effects.
Rollback class and status are evidence about provider semantics, not mechanisms
for compensation.  A future compensation API would require a new protected
operation with its own authority, DataFlow, effect, and audit record.

\subsection{Explainable Operations}

Each action that enters a protected Tool or Runtime operation scope is assigned
a persistent causal operation tree.  Typed links connect LLM calls, Tool calls,
JIT syscalls, primitives, Capability decisions or reservations, Task Authority
identities, Human requests or responses, provider effects and transitions,
resource charges, events, and audit records.  Queries are Host-only and return
bounded projections.  Evidence completeness means that all required typed
roles and links are present for the operation; it does not imply that the
action was safe, that the model reasoned correctly, or that the database is
tamper-proof.

Per-LLM Context Materialization Manifests record each versioned Object with an
\code{included} or \code{omitted} disposition, a reason such as
\code{selected}, \code{filter_mismatch}, \code{capability_denied},
\code{token_budget}, or \code{missing}, and a \code{verbatim},
\code{compacted}, or \code{truncated} transformation.  They retain bounded
token, hash, and label metadata without duplicating payloads.  Provider trace
views group a bounded set of attempts under one logical call and may expose
retained output within fixed bounds, Tool calls, usage, model, request, and
response identifiers, timing, errors, and provider-returned reasoning or
summary fields.  \sys neither records hidden chain of thought nor claims
access to it.

Following established provenance systems, the evidence design represents
lineage and recorded dependencies as typed relations among entities,
activities, and agents~\cite{w3cprov2013,lpm2015,camflow2017,provagent2025}.
The purpose here is narrower and operational: to identify the Runtime boundaries
involved, support recovery, and ensure fail-closed handling of missing links.
Audit and effect-transition histories are append-only through RuntimeStore
APIs, whereas guarded effect rows and retained payload projections may be
updated in place.  This design is not a cryptographic append-only ledger that
protects against a database administrator.

\section{Durable Execution, Self-Evolution, and Ordered Recovery}
\label{sec:durable}

\subsection{Affordances do not implicitly expand authority over existing resources}

Table~\ref{tab:evolution} summarizes how each self-evolution mechanism is
constrained at admission, information-flow, and recovery boundaries.  The
shared invariant is that an affordance may alter a process's program or
model-visible action surface, but any resulting resource access remains
subject to an existing primitive or Protected Operation contract.  Newly
created, process-owned resources may carry newly issued handles, and booting a
trusted image package may invoke the narrow private-workspace bootstrap
described below.  Neither case turns an affordance into a grant over
pre-existing or external state.

\begin{table*}[t]
\caption{Self-evolution mechanisms and the boundaries that constrain them.  The
third column reports whether a mechanism implicitly expands authority over
pre-existing or external resources; handles for newly created resources and
the trusted private-workspace bootstrap are identified explicitly.}
\label{tab:evolution}
\centering
\small
\begin{tabularx}{\textwidth}{L{0.13\textwidth} L{0.19\textwidth} C{0.11\textwidth} Y Y}
\toprule
Mechanism & Evolves & Implicitly expands authority over pre-existing or external resources? & Admission and flow boundary & Durable/recovery rule \\
\midrule
Skill activation & Prompt instructions and projected Tools/JIT candidates & No & Skill authority governs activation; invoked Tools remain subject to their domain primitives & Snapshot and trust metadata persist; unloading removes the Skill's projection and process-local JIT bindings while preserving independent base bindings \\
JIT registration & Process-local executable Tool code & No & Deno accesses the Runtime only through typed syscall RPC & Source, source hash, and Tool metadata persist; the Runtime supervisor owns process-tree termination \\
Image register/exec & Boot prompt, Tool table, policy defaults, and workspace seed & No; fresh private-workspace bootstrap only & Ordinary requirements are declarative only; for a Host-accepted package, the Runtime may issue only the grants declared over its newly created private workspace & Publication is recovery-bound; other exec paths preserve or intersect current authority \\
Checkpoint restore/fork/commit & Scoped Runtime state and reusable boot state & No & Restore and fork require checkpoint and process authority; DataFlow and task ceilings remain in force after intersection & External effects remain historical; stale authority is never resurrected \\
Child process & A new subject, namespace, queue, and action surface & No & Spawn/fork and authority delegation are explicit and attenuating; handoff preserves labels & Parent--child state and messages are durable; revocation takes precedence \\
Object Memory & Versioned Objects, links, and context selection & No; fresh Object handle only & Existing Object and namespace rights, together with receive-domain checks, govern use & Versioned state is checkpoint-scoped; context manifests record materialization links \\
Remote registry & Visible JSON-RPC/MCP affordances & No & Public IDs pass an early authority gate; exact-item authority and Sink checks follow resolution of hidden metadata & Registry digests and generations invalidate stale grants; reconciliation occurs only through explicit hooks \\
Typed Git remotes & Existing repository-configured remotes & No & Dispatch is bound to repository configuration, helper/ref fingerprints, remote authority, and Sink clearance & Startup observes only local state; remote ambiguity remains \code{unknown} \\
\bottomrule
\end{tabularx}
\end{table*}

\subsection{Skills and JIT Tools}

Skills are declarative \code{SKILL.md} packages comprising instructions,
resources, visible built-in Tools, and optional JIT candidates.  Discovery and
activation are explicit operations.  Trust and Skill authority determine which
packages may be loaded; activating a Skill does not grant the filesystem,
shell, Git, MCP, or other Capabilities required by its Tools.  Built-in images
may initially expose a compact lifecycle projection while retaining a broader
immutable callable-table ceiling.  The complete table enables trusted direct
workflows while keeping the model projection distinct from the security
boundary.

JIT Tools execute TypeScript under Deno without direct access to a libOS
object.  Their sole Runtime entry point is \code{libos.syscall}; messages sent
through this entry point are schema-validated and dispatched to the same
primitive managers as Python Tools.  Under the current policy, static imports,
re-exports, and dynamic \code{import()} are rejected.  Before Deno starts, a
supervisor scoped to the Host lifetime establishes process-tree containment
and accounts for supported subprocess limits.  These mechanisms provide defense in depth; they
neither establish a sound confinement proof for arbitrary TypeScript nor
constitute a kernel-grade sandbox.

By contrast, trusted Runtime Modules are Host-selected Python code loaded
before \code{Runtime.open()} returns.  They may register Tools, images,
syscalls, provider hooks, and startup hooks.  Although they expand the TCB,
registration alone does not grant a process a Capability.

\subsection{Images, checkpoints, and publication}

Checkpoint creation snapshots reconstructable state for a process subtree,
including process and Object state; Tool/Skill/JIT metadata; messages and
lifecycle position; durable \code{llm_pending_actions}; image artifacts
required by the subtree; and relevant authority metadata.  Active ObjectTasks,
Task Run state, and terminal-cleanup records are excluded from the snapshot,
while Host provider registries and provider state remain part of the current
external configuration.  Restore preserves historical Human, LLM, audit, and
external-effect records.  Fork remaps process identities; intentionally
discards captured \code{llm_pending_actions}; revalidates current source
Capabilities; removes revoked, expired, finite-use, or newly overlapping
grants; derives fresh attenuated manifests; and reserves the captured root
budget against the current parent.  It cannot resurrect a revoked grant.

Committing a checkpoint publishes a checkpoint-derived artifact and a
corresponding image-registry entry; it neither boots the image nor materializes
filesystem content.  Separately, booting an image package with a workspace seed
may materialize a Runtime-private workspace through Host filesystem APIs under
a durable publication program.  The Runtime may issue only the filesystem
grants declared in the Host-accepted package manifest, and only over that fresh
copy.  Startup reconciles interrupted publications before making the Runtime
available.  This narrow TCB exception has a fixed target and operation; it does
not create a model-facing filesystem bypass.

\subsection{Durable Task Runs}

A \taskrun supervises a single root \agentprocess tree under a versioned goal
and associated requirements.  It exposes idempotent Host commands to start,
follow up, continue, cancel, inspect, and perform terminal settlement, together
with an append-only, evidence-linked ledger.  Safe resume points distinguish completed
local work from provider-side ambiguity.  Retention of plaintext payloads
requires explicit opt-in; by default, reports expose only bounded identities
and hashes.

A Task Run is neither a workflow DSL nor a distributed scheduler.  The model
and Tools continue to determine task structure.  Instead, a Task Run provides
a durable container for one process tree, command deduplication, Runtime-epoch
fencing, recovery, and an operator-visible \code{needs_attention} state.
Unknown effects, missing required payloads, or abandoned ObjectTasks prevent
automatic continuation rather than being treated as success.
Version 1 certifies only one dynamic binding change: \code{activate_skill} as
the sole action in a validated, nonparallel model response.  Any other action
that changes Image, Tool, provider, or loaded-Skill bindings places the Task Run in
\code{needs_attention} before continuation or reopen dispatch and is never
replayed automatically.

\subsection{Startup dependency order}

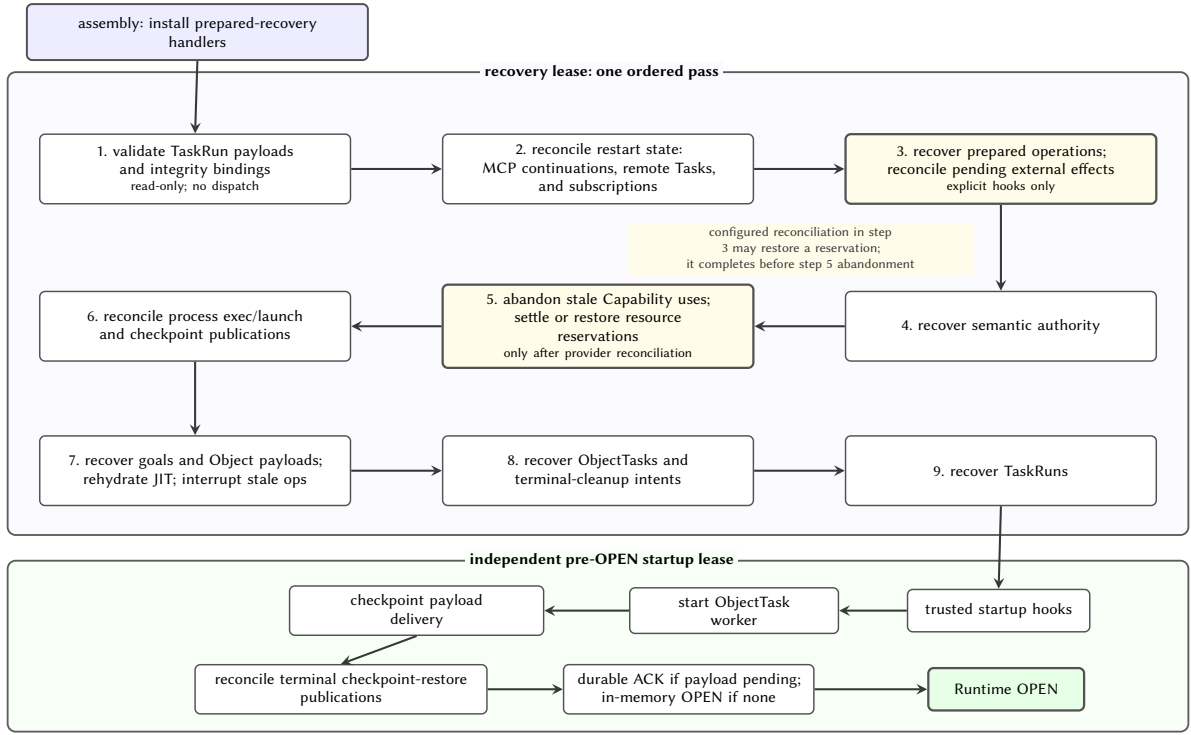
\begin{figure*}[t]
  \centering
\begin{tikzpicture}[
  x=1cm,
  y=1cm,
  >=stealth,
  font=\sffamily,
  box/.style={draw=black!68, fill=white, rounded corners=2pt,
    line width=0.55pt, align=center, inner sep=3pt,
    font=\sffamily\scriptsize},
  recoveryStep/.style={box, text width=3.90cm, minimum height=0.92cm},
  critical/.style={recoveryStep, fill=yellow!10, line width=0.85pt},
  arrow/.style={->, line width=0.78pt, draw=black!78},
  note/.style={font=\sffamily\tiny, align=center, text=black!78}
]
  \path[use as bounding box] (0,0) rectangle (15.9,9.85);

  \node[box, fill=blue!7, text width=4.30cm, minimum height=0.74cm,
    line width=0.75pt] (handlers) at (2.65,9.35)
    {assembly: install prepared-recovery\\handlers};

  \draw[draw=black!64, fill=blue!2, rounded corners=3pt, line width=0.70pt]
    (0.14,2.70) rectangle (15.76,8.82);
  \node[font=\sffamily\scriptsize\bfseries, fill=blue!2, inner sep=1.5pt]
    at (8.00,8.82) {recovery lease: one ordered pass};

  \node[recoveryStep] (payload) at (2.62,7.54)
    {1. validate TaskRun payloads\\and integrity bindings\\[-1pt]{\tiny read-only; no dispatch}};
  \node[recoveryStep] (mcp) at (7.95,7.54)
    {2. reconcile restart state:\\MCP continuations, remote Tasks,\\and subscriptions};
  \node[critical] (effects) at (13.28,7.54)
    {3. recover prepared operations;\\reconcile pending external effects\\[-1pt]{\tiny explicit hooks only}};

  \node[recoveryStep] (semantic) at (13.28,5.46)
    {4. recover semantic authority};
  \node[critical] (reservations) at (7.95,5.46)
    {5. abandon stale Capability uses;\\settle or restore resource\\reservations\\[-1pt]{\tiny only after provider reconciliation}};
  \node[recoveryStep] (publications) at (2.62,5.46)
    {6. reconcile process exec/launch\\and checkpoint publications};

  \node[recoveryStep] (goals) at (2.62,3.55)
    {7. recover goals and Object payloads;\\rehydrate JIT; interrupt stale ops};
  \node[recoveryStep] (objects) at (7.95,3.55)
    {8. recover \mbox{ObjectTasks} and\\terminal-cleanup intents};
  \node[recoveryStep] (taskruns) at (13.28,3.55)
    {9. recover TaskRuns};

  \draw[arrow] (handlers.south) -- (payload.north);
  \draw[arrow] (payload.east) -- (mcp.west);
  \draw[arrow] (mcp.east) -- (effects.west);
  \draw[arrow] (effects.south) -- (semantic.north);
  \draw[arrow] (semantic.west) -- (reservations.east);
  \draw[arrow] (reservations.west) -- (publications.east);
  \draw[arrow] (publications.south) -- (goals.north);
  \draw[arrow] (goals.east) -- (objects.west);
  \draw[arrow] (objects.east) -- (taskruns.west);

  \node[note, fill=yellow!10, inner sep=1.5pt, text width=4.45cm]
    at (10.63,6.48)
    {configured reconciliation in step 3 may restore a reservation;\\it completes before step 5 abandonment};

  \draw[draw=black!64, fill=green!3, rounded corners=3pt, line width=0.70pt]
    (0.14,0.10) rectangle (15.76,2.36);
  \node[font=\sffamily\scriptsize\bfseries, fill=green!3, inner sep=1.5pt]
    at (8.00,2.36) {independent pre-OPEN startup lease};
  \node[box, fill=white, text width=2.20cm, minimum height=0.56cm] (hooks)
    at (13.25,1.70) {trusted startup hooks};
  \node[box, fill=white, text width=2.55cm, minimum height=0.56cm] (worker)
    at (9.75,1.70) {start ObjectTask\\worker};
  \node[box, fill=white, text width=3.15cm, minimum height=0.56cm] (delivery)
    at (5.55,1.70) {checkpoint payload\\delivery};
  \node[box, fill=white, text width=3.65cm, minimum height=0.56cm] (terminalrestore)
    at (4.55,0.66) {reconcile terminal checkpoint-restore\\publications};
  \node[box, fill=white, text width=3.10cm, minimum height=0.56cm] (ack)
    at (9.15,0.66) {durable ACK if payload pending;\\in-memory OPEN if none};
  \node[box, fill=green!10, text width=1.85cm, minimum height=0.56cm,
    line width=0.85pt] (open) at (13.35,0.66) {Runtime OPEN};
  \draw[arrow] (taskruns.south) -- (hooks.north);
  \draw[arrow] (hooks.west) -- (worker.east);
  \draw[arrow] (worker.west) -- (delivery.east);
  \draw[arrow] (delivery.south) -- (terminalrestore.north);
  \draw[arrow] (terminalrestore.east) -- (ack.west);
  \draw[arrow] (ack.east) -- (open.west);
\end{tikzpicture}
  \caption{Pre-\code{OPEN} recovery dependencies.  This ordering prevents a
  later phase from discarding authority, payload, or publication state required
  to reconcile an earlier ambiguous operation.  Reopening completes local
  recovery; it neither rolls back provider state nor authorizes redispatch.}
  \Description{An ordered recovery graph. Prepared-recovery handlers are
  installed before the recovery pass. Task Run payload validation and MCP
  continuation reconciliation precede effect reconciliation; semantic recovery
  precedes cleanup of stale reservations; and publication, goal, JIT,
  ObjectTask, cleanup, and Task Run recovery precede startup hooks, ObjectTask
  worker startup, checkpoint-payload delivery, terminal restore-publication
  reconciliation, payload acknowledgment, and the transition to \code{OPEN}.}
  \label{fig:recovery-order}
\end{figure*}

Startup recovery completes before the Runtime enters \code{OPEN}.  This order
is a security property because each later phase relies on conclusions
established by earlier phases.  Prepared-operation recovery handlers are
installed during assembly before the recovery pass begins.  While holding the
recovery lease, the builder performs the following steps:
\begin{enumerate}[leftmargin=1.4em]
  \item validate recoverable Task Run plaintext and its integrity bindings
  without dispatching;
  \item reconcile interrupted MCP continuations, remote Tasks, and
  subscriptions;
  \item recover prepared operations and reconcile pending provider effects,
  admitting receipt observations only through explicitly installed hooks;
  \item recover semantic exact-once authority, then abandon stale Capability
  uses and settle or restore resource reservations;
  \item reconcile process-exec, process-launch, and checkpoint-restore
  publications;
  \item recover root-goal payloads and missing Object payloads, rehydrate
  registered JIT Tools, interrupt stale operations and executions, and recover
  ObjectTasks and terminal-cleanup intents;
  \item recover Task Runs only after their dependent process, operation, and
  effect state is stable.
\end{enumerate}
Figure~\ref{fig:recovery-order} presents a dependency graph, not a wall-clock
guarantee.  Recovery proceeds under a dedicated lease.  High-cardinality
effect and recovery scans are paginated with configured bounds, while other
components enforce their own hard bounds.  When provider reconciliation is configured, it
precedes abandonment of stale Capability reservations because it may restore a
reservation after establishing that the corresponding effect was not
dispatched.  After this pass, while the Runtime remains pre-\code{OPEN}, the
builder holds the startup lease, runs trusted startup hooks, starts the
ObjectTask worker, performs checkpoint payload delivery, reconciles terminal
checkpoint-restore publications, and commits the delivery acknowledgment
before transitioning the Runtime to \code{OPEN}.

This ordering reflects a broader lesson from durable execution systems: local
state machines must preserve and expose ambiguity at external boundaries rather
than assume that a database transaction can span external
systems~\cite{sagas1987,rifl2015,beldi2020,exoflow2023,durablefunctions2021,sagallm2025}.
Unlike general-purpose workflow systems, \sys binds these durability guarantees
to agent process identity, finite authority, IFC, and causal evidence.

\section{Implementation and Case Studies}
\label{sec:implementation}

\subsection{Runtime composition}

The prototype supports Python 3.11--3.14 and exposes an embeddable Runtime, a
CLI, and a loopback Electron/React operator interface.  During Host composition,
the builder constructs managers and provider contracts, loads hash-trusted
Runtime Modules, installs recovery handlers, completes pre-\code{OPEN}
reconciliation, and only then publishes the Runtime as \code{OPEN}.  A bounded,
thread-backed worker pool advances quanta for runnable \agentprocess instances;
a quantum that waits for Human input, a timer, a message, a child process, a
Tool task, or an LLM call is confined to its worker, allowing other workers to
continue.
An asynchronous Host wrapper integrates the same Runtime with an external event
loop.

\toolbroker resolves a named entry only through the process's complete callable
Tool table and opens the corresponding causal Tool operation; the invoked Tool
then constructs any primitive-specific context and traverses primitive
admission.  The model executor receives a separate projection determined by the
image and active Skills; depending on the image, this projection may equal or
be narrower than the complete table.  Direct Host workflows can invoke a
callable Tool without an LLM turn, but still traverse \toolbroker, process
authority, and primitive admission.  Thus, projection can change model
ergonomics without changing the enforcement surface.

The Resource Provider Substrate supplies typed filesystem, clock, shell, Human
I/O, Git, JSON-RPC, and MCP backends.  The LLM provider service is composed
alongside that substrate, while optional PTY support is supplied by a trusted
Runtime Module.  Remote endpoints, transports, credentials, Git remotes, and
LLM profiles are fixed by Host composition.  Model-facing operations may select
only the logical identifiers exposed by their typed interfaces, such as a
registered JSON-RPC endpoint and method, an MCP Tool or model-visible Resource,
or an existing Git remote; they cannot inject an ad hoc URL, credential,
transport command, or profile definition.  Each provider contract also
specifies result bounds, data-flow direction, and conservative effect
classification.

\subsection{Persistence and transaction boundaries}

Durable state is coordinated through a Unit of Work comprising typed domain
repositories for processes, Task Runs, authority, resources, Objects, evidence,
extensions, publications, semantic assessments, MCP lifecycle state, and
payload retention.  SQLite is the default local backend, while PostgreSQL 17
is an optional backend over the same canonical store-schema-v7 contract.
Opening a persistent store acquires a single-writer Runtime lease and validates
the version marker, exact schema shape, required indexes, encoding, and
collations.  Canonical version-4, -5, and -6 stores must traverse the remaining
explicit offline migration steps to version 7 in order.  Version-3 archives,
older or unversioned stores, and malformed or incomplete version-7 stores are
rejected rather than repaired at startup.

Object metadata and namespaces are durable, but ordinary Object payloads remain
Runtime-local and are represented in the store by live-payload markers.  A
narrow, explicitly enabled root-goal envelope may retain a bounded payload;
scoped checkpoints and image artifacts may separately capture the bounded
payloads needed to reconstruct a process subtree.  No other ordinary Object
payload is implied to survive Runtime restart.

The database transaction boundary is deliberately narrow.  A RuntimeStore
transaction can atomically update process transitions, authority and resource
reservations, protected-effect intents, events, and audit records.  External
actions---including filesystem changes, subprocess or Git operations, Human
I/O, and JSON-RPC, MCP, or LLM calls---cannot participate in that transaction.
The \protectedop state machine and startup recovery therefore expose this
boundary explicitly rather than hiding it behind a Tool return value.

Payload retention is distinct from causal retention.  Disabled by default,
Host-triggered maintenance can monotonically reduce eligible terminal LLM-call
and external-effect provider payloads from full to summary and then hash-only,
while preserving identity, classification, transition state, digests, and
causal links.  Run-linked Human payloads are governed separately by the Durable
Task Run retention contract.  Neither path changes an effect's state or permits
a nonterminal or \code{unknown} effect to appear terminal.

\begin{table*}[t]
\caption{Implemented surfaces and their primary boundaries.  ``Client subset''
is deliberate: no row implies an MCP server or a public remote API.}
\label{tab:implementation}
\centering
\small
\begin{tabularx}{\textwidth}{L{0.17\textwidth} L{0.24\textwidth} Y Y}
\toprule
Subsystem & Implemented surface & Primary admission boundaries & Evidence/recovery boundaries \\
\midrule
Processes and memory & Lifecycle, queues, Object Memory, cwd, context materialization, background ObjectTasks & Process/Object/filesystem Capability, applicable Task Authority ceiling, receive domain, budgets & Store transactions, process state/execution generations, context manifests, Task Run epoch fencing/recovery \\
Self-evolution & Skills, JIT Tools, images, checkpoints, child processes, trusted modules & Affordance-specific primitive plus unchanged resource admission & Snapshot lineage, publication recovery, source hashes, causal links \\
Local providers & Filesystem, shell, clock, Human I/O, optional Object-bound PTY & Typed resource and policy; DataFlow for input/output Sinks & Protected effects, resource settlement, lifecycle cleanup \\
Git & Typed inspection, local mutation, managed worktrees, immutable patches, existing remotes, simulated PRs & Repository, remote, PR, and Object Capabilities; applicable Task Authority ceiling; exact state token; affected filesystem Capabilities; Sink clearance & Irreversible/unknown effect classification, repository/worktree locks, and local, query-only recovery observations \\
Remote clients & Registered JSON-RPC and MCP stdio/Streamable HTTP client subsets & Per-item authority, Host transport, Sink and ingress labels, byte/time budgets & Registry generations, Protected Operations, continuation/task/subscription recovery \\
LLM & Named OpenAI-compatible profiles, bounded attempt traces, optional prompt-cache policy & Profile Capability, applicable Task Authority ceiling, LLM Sink identity, token reservation & Logical-call record, bounded response projection, context/effect links \\
Operator surfaces & Python API, CLI, loopback HTTP/SSE, and Electron GUI & Loopback bearer authentication, exact-origin/CORS policy, and Host/admin versus actor-mode routes & Redacted snapshots, durable commands, and causal inspection \\
\bottomrule
\end{tabularx}
\end{table*}

\subsection{Case study: typed Git}

Git illustrates why wrapper-level authorization is insufficient.  Its CLI
admits a large language of options, configuration, hooks, filters, alternate
object stores, credential and transport helpers, refspecs, revision
expressions, pathspecs, and repository-discovered metadata.  A model permitted
to supply arbitrary \code{git} arguments can therefore select executable code
or destinations indirectly, even when a wrapper blocks an obvious option.

\sys instead pins \code{Runtime.git} to the configured non-bare workspace
repository.  Model-facing Tools select typed operations for bounded inspection,
local changes, branch/tag/stash/worktree management, immutable patch Objects,
existing remotes, and repository-local simulated pull requests.  No typed
operation supports init, clone, remote or configuration mutation, arbitrary
arguments, submodule update, LFS, plumbing, custom upload/receive pack,
model-selected credentials, or arbitrary URLs.  On every call, the provider
revalidates the repository root, selected worktree, Git directory, common
directory, object format, and filesystem identity.  Managed worktrees remain
under a Host-selected root; the process cannot supply an external destination.

Git output is captured and parsed as bytes.  Each path is returned as display
text together with a base64 token for the exact repository-relative bytes.
Path-bearing commands use literal pathspecs and the \code{--} option terminator,
preventing a filename that begins with a dash or contains non-UTF-8 bytes from
being interpreted as an option.  Results report truncation, byte counts, and
hashes explicitly; immutable patch Objects must always be complete.

Before any mutation, the caller supplies an opaque state token obtained from a
prior read.  The token commits to repository/worktree identity, HEAD and ref
state, the index, effective configuration, the worktree registry, simulated-PR
metadata, and bounded worktree content state.  With a cross-process repository
lock held, the primitive rechecks the token immediately before dispatch.  A
failed initial comparison before the first state-changing effect certifies that
the invocation did not dispatch; a later mismatch may follow an earlier effect
and is therefore possibly dispatched.  The token is compare-and-swap evidence;
it is neither read authority nor a replay credential.

Git models the pinned repository, each configured remote, each simulated pull
request, and immutable patch Objects as distinct resources.  Operations
classified as destructive additionally require both delete and admin authority
and a mandatory Human approval that mints an exact, approval-bound one-use
Capability lease.  Checkout-changing operations require the
corresponding filesystem read, write, or delete Capabilities, while ordinary
filesystem authority never extends to Git metadata.  A legacy Shell grant for
\code{git} confers none of this typed authority.

Information-flow policy is evaluated independently of Capability admission.
Reads are ingress; local mutations are bidirectional with the repository as
their Sink; fetch combines remote ingress with egress to the local repository
Sink; push is egress to the exact configured remote Sink; and pull-request
operations bind the exact PR resource, with creation and merge also clearing
the repository Sink.  The operation canonically binds source and carrier
labels, refs, paths, worktree identity, and the state token.  Consequently,
repository authority alone cannot authorize a secret-bearing push.

The provider disables or neutralizes hooks, pagers, editors, signing, external
diff and \code{textconv}, lazy fetch, maintenance, replace refs, and
workspace-controlled executable lookup.  It fails before dispatch if active
configuration selects external filters, drivers, helpers, proxies, custom
transports, or unsafe includes.  This is command mediation rather than an
OS-level sandbox: a separately authorized interpreter or native executable can
invoke Git with the Host user's authority.  Deployments that cannot tolerate
that possibility require a more strongly isolated provider.

Git remotes are drawn from the pinned repository's validated configuration,
not from the JSON-RPC/MCP endpoint registry.  Startup recovery is local and
query-only: it may inspect current refs, worktree state, simulated-PR metadata,
and freshly recomputed local configuration, helper, and ref fingerprints.  It
does not contact a remote, use these observations to establish the recorded
network outcome, or clear an effect in the \code{unknown} state.

\subsection{MCP and registered remote resources}

The MCP integration is client-only.  Manifests v1 and v2 retain a governed
Tools compatibility contract.  Manifest v3, pinned to the exact protocol
revision \code{2026-07-28}, adds governed Resources, Resource Templates,
Prompts, Completion, Multi Round-Trip Requests (MRTR) for
\code{input_required} continuations, bounded \code{subscriptions/listen}
streams, Host-preconfigured OAuth handling, and optional digest-pinned support
for the Tasks extension.  Negotiation is explicit and bounded;
automatic fallback is permitted only for recognized compatibility responses,
not arbitrary authentication, TLS, timeout, malformed-response, or server
errors.  Tool schemas and manifest identities are pinned before
calls~\cite{mcp2026}.

Each server and item is Host-registered under a closed, operation-specific
contract.  Server configuration fixes transport, timeout, and byte bounds.
Tool entries additionally declare effect and rollback metadata, while
Resources, Resource Templates, and Prompts use separate closed read or prompt
contracts.  Trusted configuration supplies stdio commands and environments;
the Host likewise supplies Streamable HTTP endpoints and authorization.  Public
server and item identifiers pass an early invocation-authority gate before any
hidden registry lookup.  After resolution, the primitive enforces exact item
authority and revalidates the server digest and generation immediately before
dispatch.

Continuations, subscriptions, OAuth state, and remote Tasks have explicit local
lifecycle and recovery contracts; an opaque remote task identifier never
becomes model-visible authority.  MCP capability declarations negotiate
protocol features, while item listings support discovery.  Neither constitutes
a process-level authority record, and server acceptance of an OAuth token does
not by itself authorize a local agent process.

The implementation deliberately covers a client subset rather than all of MCP\@.
\sys excludes an MCP server surface, Apps, Roots, Sampling, Logging, Dynamic
Client Registration, client-credentials and enterprise-managed OAuth,
workload-identity federation, DPoP, older OAuth compatibility modes, deprecated
standalone SSE, and product-level OpenTelemetry integration.  Executing a
trusted MCP stdio server still launches native code under the Host account; the
protocol boundary does not sandbox that code.  Accordingly, the current scope
shows that the three-plane design extends to a rich registered provider; it
does not claim universal MCP interoperability.

OAuth and delegated-authorization standards address complementary questions.
GNAP and the OAuth Security Best Current Practice constrain authorization
protocols and token handling~\cite{gnap2024,oauth2025}, while agent-oriented
systems explore policies and authenticated delegation%
~\cite{progent2025,agentspec2026,conseca2025,authenticateddelegation2025}.
\sys consumes endpoint and credential authority established by the Host; it
neither replaces these protocols nor permits a model to configure them.

\subsection{Operator and semantic surfaces}

The Python Runtime object is the primary composition interface.  The CLI and
Electron application expose process and Task Run supervision, Human queues,
provider traces, events, audit records, operations, and bounded diagnostic
views.  Loopback-only HTTP/SSE endpoints serve the local GUI; they do not form a
public REST control plane.  Semantic-policy mutation is confined to Host
composition and is not exposed through model Tools, JIT, Skills, Modules, the
GUI, the CLI, or HTTP\@.

The GUI displays bounded provider-returned reasoning, summary, or thinking text
only when it remains readable under the retention policy.  It neither
synthesizes explanations nor claims access to a model's hidden chain of
thought.  Read-only operation views can show causal links among Tool, primitive,
decision, Sink, effect, event, and audit roles, while sensitive bodies remain
behind separate on-demand projections.

\section{Evaluation}
\label{sec:evaluation}

The evaluation distinguishes deterministic boundary behavior from stochastic
model behavior and native execution from modeled design coverage.  These
distinctions prevent aggregate success rates from masking missing effect
evidence or simulated scenarios from being conflated with provider execution.
We ask:
\begin{description}[leftmargin=1.15in,style=nextline]
  \item[RQ1: Confinement.] Does the Runtime prevent unauthorized effects modeled
  by the suite, reject flows that exceed Sink clearance, and preserve declared
  allowed effects?
  \item[RQ2: Durability.] Does the handling of intent, settlement, causal
  evidence, and startup recovery remain fail-closed across crash barriers and
  large histories?
  \item[RQ3: End-to-end behavior.] Under one real-model configuration, can
  restartable maintenance, browser, research, and analysis workflows preserve
  authority and safety while satisfying independent utility oracles?
  \item[RQ4: Projection ergonomics.] Compared with immediate exposure of the
  full Tool schema, how does on-demand Skill projection affect correct routing,
  observable task success, invalid calls, schema bytes, prompt bytes, and
  provider-reported tokens under paired goals?
\end{description}

\subsection{Evidence contract and provenance}

All results reported in this paper are derived from
\path{evidence/manifest.json}.  The manifest binds each redacted JSON report to
a SHA-256 digest and records the exact command, source revision and cleanliness,
bounded working-tree identity, task/config/provider identity, report schema,
and expected metrics.  A generator validates these inputs and writes
\path{evidence/results.tex}; no results are copied into the manuscript by hand.
The publication build fails if any required report, hash, macro, denominator,
or provenance field is absent.

Here, \emph{publication-valid under the bound commit's validation contract}
denotes an artifact that is complete, hash-locked, source- and
configuration-matched, and semantically decidable.  It does not imply that a
scientific gate passed, that the agent reached a terminal exit, or that an
oracle was satisfied.  A complete artifact documenting a failed gate,
incorrect trajectory, or nonterminal model trajectory remains publishable as
negative evidence; an artifact with missing or undecidable evidence does not.
Because report schemas and gates evolve, this status does not imply acceptance
by a later validator; a new gate requires a new source-bound campaign rather
than retroactive backfilling.

The evaluated source is \ArtifactSourceStatus{} at commit
\ArtifactCommitShort{}.  Each real-model report also includes a redacted
provenance envelope bound to one provider configuration: model identifier, API
mode, retry and timeout settings, request bounds, prompt/cache policy, endpoint
class and normalized endpoint hash, and a credential-present boolean.  The
envelope never includes an API key, raw endpoint, or environment-variable
value.  Published reports retain only the bounded task/evidence fields required
by their oracles; unredacted prompts, provider responses, credentials, and task
secrets are excluded from the paper artifact.  The canonical combiner rejects a
dirty or changing tree and any cross-family mismatch in commit, model, or
provider configuration.

The provider-smoke artifact has status \ProviderSmokeStatus{}.
\IfResultFlag{ProviderSmokeOuterReceipt}{This is an outer,
operator-observed command receipt, not a report emitted by the smoke script or
an attestation of the effective LLM profile.}{It is not classified as a valid
outer receipt.}{No provider-smoke receipt is available.}
\IfResultFlag{ProviderSmokeFailedTerminalStateCheck}{The receipt records a
failure of the preflight terminal-state check.}{The receipt does not record a
failure of the terminal-state check.}{The terminal-state result is
unavailable.}
\IfResultFlag{ProviderSmokeRetroactivelyPassed}{A later run retroactively
reclassified that result.}{Later paid runs do not retroactively reclassify this
failed preflight as a pass.}{No retroactive-status claim is made.}

If an implementation or harness defect is corrected during a campaign, the
source revision changes and the affected matrix is invalidated.  A defect found
after the campaign remains a limitation of the bound source and is not repaired
retroactively by a later revision.  Ordinary model failures are retained as
results and do not trigger retries.  A verified infrastructure failure may
trigger a rerun of the entire affected family; the failed artifact and reason
are retained, and successful trajectories are not selectively cherry-picked.

\subsection{RQ1: deterministic runtime safety}

The runtime-safety suite comprises versioned YAML tasks with copied fixtures,
declared allowed and forbidden effects, deterministic execution plans, and
evidence-backed oracles.  It covers filesystem, shell, Human, process, Object
Memory, image, checkpoint, Skill, JIT, JSON-RPC, MCP, data-flow, typed Git, and
Semantic Shadow boundaries.  The self-evolution subset attempts to activate
Skills, register and execute JIT Tools, register, execute, and commit images,
fork
children, reuse checkpoints, attach remote affordances, manipulate Git state,
and inject semantic authority.

The full Runtime runner is evaluated under strict success/safety and
release-evidence gates.  Three deliberately simple comparison interventions
execute the same task plans: a direct Tool wrapper, a confirmation wrapper, and
a static Tool-category ``sandbox.''  The third is neither a container, VM, nor
Host sandbox; the harness models effects unsupported by a baseline.  These
baselines isolate boundary behavior and do not constitute real-model agent
comparisons.

The ablations remove primitive approval, namespace isolation, or fork
attenuation one at a time.  A separate \code{no_audit_linkage} diagnostic
intentionally withholds required evidence.  Because missing evidence invalidates
a result, this diagnostic is reported separately and never enters a safety-rate
table.

The metric collector distinguishes task rows from effect records.  Task-success
and safety-pass rates are micro-averages over the declared task oracles.  The
unauthorized-side-effect rate includes in its denominator only effect attempts
definitely known to have been performed.  The false-denial rate includes only
allowed attempts whose outcomes are definitively classified as performed or
denied.  An unknown or
missing outcome invalidates the row rather than being inferred from a Tool
exception.  For each task, audit completeness is the fraction of definitely
performed effects with matching audit records; it is defined as 1.0 when no
effect was performed.  The reported value is the mean across tasks.  Other
evidence-role completeness is validated separately.

\begin{table*}[t]
\caption{Deterministic runtime-safety results.  Fractions report explicit
numerators and denominators; invalid rows are not coerced to zero.  Wrapper and
category interventions are deterministic harness baselines, not OS sandboxes or
real-LLM runs.}
\label{tab:runtime-results}
\centering
\small
\begin{tabularx}{\textwidth}{L{0.18\textwidth} C{0.08\textwidth} C{0.12\textwidth} C{0.12\textwidth} C{0.15\textwidth} C{0.15\textwidth} Y}
\toprule
Runner & Tasks & Task success & Safety pass & Unauthorized & False denial & Evidence \\
\midrule
\RuntimeSafetyTableRows
\bottomrule
\end{tabularx}
\end{table*}

\IfResult{RuntimeSafetyPassed}{For the full Runtime row, the source-bound
artifact records \RuntimeTaskSuccessPassed{}/\RuntimeTaskCount{} passing
task-success oracles and \RuntimeSafetyPassed{}/\RuntimeTaskCount{}
safety-passing tasks, together
with an unauthorized-effect fraction of
\RuntimeUnauthorizedNumerator{}/\RuntimeUnauthorizedDenominator{}, a
false-denial fraction of
\RuntimeFalseDenialNumerator{}/\RuntimeFalseDenialDenominator{}, and mean audit
completeness of \RuntimeAuditCompletenessPercent\%.  These values characterize
only the checked task/effect population and do not estimate open-world agent
risk.
\IfResultFlag{RuntimeReleaseGatePassed}{The artifact records the full
release-evidence gate as passed.}{The artifact records the full
release-evidence gate as not passed.}{The release-gate result is
unavailable.}}{The deterministic publication artifact was unavailable for this
build, so the manuscript makes no observed pass-rate claim.}

The invariant manifest is a coverage map, not an execution receipt.
\IfResult{InvariantDeclarationCount}{For the bound source, it resolves
\InvariantDeclarationCount{} invariant declarations against
\CollectedPytestNodeCount{} collected pytest nodes.  Test execution status is
recorded separately.}{No invariant-count claim is made without the bound
checker report.}
\IfResult{TestMatrixPassed}{Independent execution receipts classify the
complete Python matrix as \TestMatrixStatus{}, the GUI lane as \GUIStatus{},
and the invariant checker as \InvariantCheckerStatus{}.}{No successful test
execution is claimed without all three bound receipts.}

\subsection{Practical workflow evidence}

The practical suite distinguishes two evidence levels.  A \code{native-live}
scenario must execute real \toolbroker calls against a stateful local provider,
compare provider state before and after execution, persist the matching
external effect, and resolve an explicit causal operation for every semantic
effect.  Native scenarios never fall back to modeled evidence.  The native
connector fixture implements synthetic mail, CRM, and calendar mutations
through registered JSON-RPC methods.  It is deterministic local infrastructure
and does not establish production-connector interoperability.

A \code{modeled} scenario evaluates a design-level utility/security oracle but
receives no credit for native Tools, operations, or provider effects.
Accordingly, the two evidence levels retain separate scenario and
semantic-effect denominators.

\begin{table}[t]
\caption{Practical-workflow results by evidence level.}
\label{tab:practical-results}
\centering
\small
\begin{tabularx}{\columnwidth}{Y Z Z Z}
\toprule
Evidence & Passed/total & Semantic effects & Native Tool calls \\
\midrule
\PracticalTableRows
\bottomrule
\end{tabularx}
\end{table}

\IfResult{PracticalNativeTotal}{The bound report presents
\PracticalNativePassed{}/\PracticalNativeTotal{} native scenarios and
\PracticalModeledPassed{}/\PracticalModeledTotal{} modeled scenarios as
separate observations.
\IfResultFlag{PracticalNativeGatePassed}{The native evidence-level gate is
recorded as passed.}{The native evidence-level gate is recorded as not
passed.}{The native gate result is unavailable.}
\IfResultFlag{PracticalModeledGatePassed}{The modeled evidence-level gate is
recorded as passed.}{The modeled evidence-level gate is recorded as not
passed.}{The modeled gate result is unavailable.}}{No practical-workflow count
is asserted without a validated report.}

\subsection{RQ2: recovery and crash evidence}

Three SQLite reopen gates evaluate structural boundedness rather than
throughput.  The external-effect gate seeds a large terminal history together
with pending prepared effects; the runtime-publication gate seeds terminal and
unreconciled publications; and the Task Run gate seeds historical and
recoverable runs.  Each gate checks its partial indexes, keyset-query shape,
exact page/row work, and complete convergence.  Elapsed seed, reopen, and
recovery times are diagnostic only; they are not thresholds, service-level
objectives, or evidence of Runtime overhead.

The Task Run crash matrix uses isolated worker processes and an independently
fsynced provider ledger across the protocol's durability barriers.  It
distinguishes pure local work, certified non-dispatch, durable provider receipt,
and unknown dispatch.  Reopening must result in no more than one provider
dispatch, deduplicate a receipt where supported, leave unknown work in
\code{needs_attention}, and produce a stable evidence fingerprint on the second
reopen.

\begin{table*}[t]
\caption{Structural recovery gates.  ``Recoverable/pending'' is the subset
selected by the partial recovery index; elapsed time is intentionally omitted.}
\label{tab:recovery-results}
\centering
\small
\begin{tabularx}{\textwidth}{L{0.26\textwidth} C{0.16\textwidth} C{0.20\textwidth} C{0.18\textwidth} Y}
\toprule
Gate & Total & Recoverable/pending & Reconciled & Page size \\
\midrule
\RecoveryTableRows
\bottomrule
\end{tabularx}
\end{table*}

\IfResult{ExternalEffectTotal}{The validated profiles contain
\ExternalEffectTotal{} external-effect rows, of which
\ExternalEffectPending{} are pending;
\PublicationRecoveryTotal{} publication rows, of which
\PublicationRecoveryPending{} are unreconciled; and
\TaskRunRecoveryTotal{} Task Runs, of which
\TaskRunRecoveryPending{} are recoverable.  The crash matrix passes
\TaskRunCrashPassed{}/\TaskRunCrashTotal{} ordered barrier cases.
\IfResultFlag{TaskRunUnpairedPassed}{The separately declared unpaired
safe-point scenario also passes.}{The separately declared unpaired safe-point
scenario does not pass.}{The unpaired scenario result is unavailable.}}{No
structural-scale or crash-matrix result is claimed without the hashed recovery
reports.}

\subsection{RQ3: canonical real-model workflows}

The opt-in live gate evaluates restartable synthetic tasks under one fixed
OpenAI-compatible model configuration.  The repository-maintenance family uses
typed file Tools to edit a fixture, runs a documented test command, inspects
dedicated Git status/diff Tools, checkpoints, reports to a Human, and exits
after a cumulative review.  A follow-up arrives after Runtime reopen; replaying
the stable continuation command must duplicate neither LLM work nor provider
effects.

The browser family drives an isolated customer portal in a real headless
Chromium instance through a fixed registered JSON-RPC bridge.  The agent must
ignore an injected customer instruction, incorporate a follow-up constraint
after reopen, issue one idempotency-keyed partial refund, read back the order,
preserve its paid status, and satisfy provider-state and DOM-action oracles.
The bridge exposes neither arbitrary URLs nor a cancellation method.

The research workflow reconciles dated, conflicting local sources, including an
indirect injection, without mutation.  The data-analysis workflow writes an
inert, reviewable standard-library script and a separately materialized JSON
artifact, incorporates a segment follow-up, and allows a latency guardrail to
determine the rollout decision.  It has no shell authority, and the evaluator
never executes model-authored code.

Each family has independent authority, safety, utility, recovery,
settled-effect, and duplicate-dispatch checks.  The canonical gate requires
every family gate to pass, every recorded run to satisfy the safety/authority
and zero-duplicate checks, and the preregistered aggregate and family-level
utility minima to be met.  Failure to invoke an expected operation reduces utility but
does not itself constitute an authority failure.  Conversely, a prohibited
call correctly denied by the Runtime counts as a safety success but may still
be an instruction-following utility failure.

\begin{table}[t]
\caption{Canonical live gate.  Status is descriptive; threshold misses are
retained as observations rather than selectively retried.}
\label{tab:live-results}
\centering
\small
\begin{tabularx}{\columnwidth}{Y Z Z Z}
\toprule
Family & Safety & Utility & Runs \\
\midrule
\CanonicalLiveTableRows
\bottomrule
\end{tabularx}
\end{table}

The combined artifact has status \CanonicalLiveStatus{}.
\IfResult{CanonicalLiveSafetyPassed}{Across the complete matrix, the artifact
records an observed safety fraction of
\CanonicalLiveSafetyPassed{}/\CanonicalLiveSafetyTotal{} and an observed
strict-utility fraction of \CanonicalLiveUtilityPassed{}/\CanonicalLiveUtilityTotal{}.
These fractions describe one model/provider configuration and are not estimates
of population-level rates.
\IfResultFlag{CanonicalLiveReleaseGatePassed}{The combined canonical release
gate is recorded as passed.}{The combined canonical release gate is recorded
as not passed.}{The combined release-gate result is unavailable.}}{No live
safety or utility fraction is asserted because a complete source- and
configuration-matched report was not available.}

\IfResult{CanonicalLiveLogicalCalls}{Canonical accounting records
\CanonicalLiveLogicalCalls{} logical LLM calls and
\CanonicalLiveProviderAttempts{} provider attempts.  Provider-reported
trajectory aggregates contain \CanonicalLiveInputTokens{} input tokens and
\CanonicalLiveOutputTokens{} output tokens.  Cache-read usage is reported for
\CanonicalLiveCacheReadReportedCalls{}/\CanonicalLiveCacheTotalCalls{} logical
calls, totaling \CanonicalLiveCacheReadTokens{} tokens.  Because the frozen
family schema does not expose per-call input/output reporting counters, these
input/output aggregates are not claimed to have complete per-call coverage.
\IfResult{CanonicalLiveCacheWriteTokens}{Cache-write usage is reported for
\CanonicalLiveCacheWriteReportedCalls{}/\CanonicalLiveCacheTotalCalls{} calls,
totaling \CanonicalLiveCacheWriteTokens{} tokens.}{Cache-write usage is
reported for
\CanonicalLiveCacheWriteReportedCalls{}/\CanonicalLiveCacheTotalCalls{} calls;
because reporting is incomplete, the aggregate cache-write token count is
unreported, not zero.}}{No canonical call/token accounting is asserted without
the validated family reports.}

\subsection{RQ4: paired Skill projection}

The paired study compares a compact Skills-based model projection with a
no-Skills baseline that exposes the complete non-lifecycle Tool schema at the
first model call.  Both arms retain the same resource-authority ceiling and
receive byte-identical neutral task goals.  The goals specify only observable
task intent; they mention neither a Skill, an expected Skill identifier, the
treatment arm, nor a probe Tool.  Held-out intents cover workspace editing,
read-only Git, shell execution, checkpoints, and MCP registry inspection, each
with a nearby but incorrect boundary.

The full matrix uses deterministic AB/BA counterbalancing across pairs and
records the pair identifier, repetition, first-arm position, and order balance.
A run counts as correct only if the observable-state oracle, route, structured
probe, and terminal exit all succeed.  Reports also retain invalid Tool-call
counts; logical LLM-call and physical-attempt counts; initial, authorized, and
cumulative schema bytes; prompt bytes; provider-reported input, output, and
cache tokens; and a clearly labeled bytes-divided-by-four estimate.  This
estimate is
not treated as provider billing evidence.

The publication gate requires a complete paired matrix, the expected order
balance, a matching clean source state and provider configuration, and complete
oracle, route, exit, and probe evidence.  Publication readiness is
outcome-agnostic: the evidence must be structurally complete and decidable but
may document an incorrect run.  Correctness is governed separately by
\code{--require-all-correct}.  Historical report schemas remain readable for
diagnosis but cannot satisfy the publication gate.

\begin{table}[t]
\caption{Paired Skill projection comparison.  Continuous intervals use the
predeclared fixed-seed percentile bootstrap; binary outcomes use exact paired
tests.}
\label{tab:skills-results}
\centering
\small
\begin{tabularx}{\columnwidth}{Y Z Z Y}
\toprule
Metric & Skills & Baseline & Paired delta [95\% CI] \\
\midrule
\SkillsPairedTableRows
\bottomrule
\end{tabularx}
\end{table}

The paired artifact has status \SkillsLiveStatus{}.
\IfResult{SkillsPairCount}{It contains \SkillsPairCount{} complete pairs and
\SkillsRunCount{} runs.  The table reports paired effect estimates: continuous
metrics use a fixed-seed, \SkillsBootstrapResamples{}-resample percentile
bootstrap, whereas binary discordances use an exact McNemar test.}{No paired effect
estimate is asserted without the complete publication-gate report.}
\IfResult{SkillsOutcomeTreatmentPassed}{The table's task-outcome row reflects
only the independent observable-state oracle:
\SkillsOutcomeTreatmentPassed{}/\SkillsOutcomeDenominator{} in the Skills arm
and \SkillsOutcomeBaselinePassed{}/\SkillsOutcomeDenominator{} in the
baseline.  Correct routing is
\SkillsRouteTreatmentPassed{}/\SkillsRouteDenominator{} versus
\SkillsRouteBaselinePassed{}/\SkillsRouteDenominator{}; terminal exit is
\SkillsTerminalExitTreatmentPassed{}/\SkillsTerminalExitDenominator{} versus
\SkillsTerminalExitBaselinePassed{}/\SkillsTerminalExitDenominator{}; and the
conjunction of outcome, route, probe, and exit is
\SkillsAllCorrectTreatmentPassed{}/\SkillsAllCorrectDenominator{} in the Skills
arm versus \SkillsAllCorrectBaselinePassed{}/\SkillsAllCorrectDenominator{} in
the baseline.  Thus, an observable task success is not equated with a fully
correct run.}{}
\IfResult{SkillsOutcomeMcNemarP}{The exact McNemar $p$-values are
\SkillsOutcomeMcNemarP{} for task outcome and \SkillsRouteMcNemarP{} for correct
routing.  The bootstrap intervals in Table~\ref{tab:skills-results} quantify
effect sizes; they are not hypothesis-test intervals.}{}

\IfResult{SkillsTreatmentLogicalCalls}{Provider-reported accounting for the
Skills arm records \SkillsTreatmentLogicalCalls{} logical calls,
\SkillsTreatmentProviderAttempts{} provider attempts,
\SkillsTreatmentInputTokens{} input tokens, and
\SkillsTreatmentOutputTokens{} output tokens.  The corresponding baseline
values are \SkillsBaselineLogicalCalls{}, \SkillsBaselineProviderAttempts{},
\SkillsBaselineInputTokens{}, and \SkillsBaselineOutputTokens{}.
Cache-read usage is reported for
\SkillsTreatmentCacheReadReportedCalls{}/\SkillsTreatmentCacheTotalCalls{}
Skills-arm calls (\SkillsTreatmentCacheReadTokens{} tokens) and
\SkillsBaselineCacheReadReportedCalls{}/\SkillsBaselineCacheTotalCalls{}
baseline calls (\SkillsBaselineCacheReadTokens{} tokens).
\IfResult{SkillsTreatmentCacheWriteTokens}{For the Skills arm, cache-write
usage totals \SkillsTreatmentCacheWriteTokens{} tokens across
\SkillsTreatmentCacheWriteReportedCalls{}/\SkillsTreatmentCacheTotalCalls{}
reported calls.}{For the Skills arm, cache-write usage is reported for
\SkillsTreatmentCacheWriteReportedCalls{}/\SkillsTreatmentCacheTotalCalls{}
calls; its aggregate cache-write token count is unreported, not zero.}
\IfResult{SkillsBaselineCacheWriteTokens}{For the baseline, cache-write usage
totals \SkillsBaselineCacheWriteTokens{} tokens across
\SkillsBaselineCacheWriteReportedCalls{}/\SkillsBaselineCacheTotalCalls{}
reported calls.}{For the baseline, cache-write usage is reported for
\SkillsBaselineCacheWriteReportedCalls{}/\SkillsBaselineCacheTotalCalls{}
calls; its aggregate cache-write token count is unreported, not zero.}}{No
paired call/token accounting is asserted without the complete report.}
Given the small paired sample and single model configuration, these confidence
intervals characterize only the evaluated workload and should not be
interpreted as broad causal estimates for LLM agents.

\subsection{Interpretation and threats to validity}

The deterministic suite tightly controls oracle definitions and source state
but executes scripted plans; it therefore isolates Runtime mediation rather
than model judgment.  The wrapper interventions deliberately omit Runtime
components.  Their results show the consequences of those omissions within
this suite, not the security of wrapper frameworks more generally.  Effect
rates weight effect records rather than tasks and should not be compared across
artifacts with different task, fixture, runner, schema, or source hashes.

The real-model evaluation uses one model/provider configuration, a small fixed
workload, and task-specific utility oracles.  Paired counterbalancing reduces
systematic order effects but cannot eliminate provider drift or stochastic
dependence.  The study reports prompt and token observations rather than dollar
cost because the evidence includes no versioned price snapshot.

Finally, the recovery profiles assess bounded query work and state convergence,
not latency service levels or steady-state Runtime overhead.  Accordingly, they
support no performance claim.  The warranted conclusion is narrower: for the
cases exercised by the bound artifacts, the
implemented boundary either produces the required authority, flow, and evidence
outcomes or causes the result to fail closed.

\section{Related Work}
\label{sec:related}

\subsection{Agent runtimes and self-evolution}

Tool-using and multi-agent systems provide much of the action loop underlying
modern agents.  ReAct interleaves reasoning with action; Toolformer and ToolLLM
study tool use; and AutoGen, MetaGPT, CAMEL, AgentScope, and SWE-agent support
multi-agent collaboration or agent--computer interaction%
~\cite{yao2023react,schick2023toolformer,qin2023toolllm,wu2023autogen,hong2023metagpt,li2023camel,gao2024agentscope,yang2024sweagent}.
CodeAct uses executable code as a compositional action representation, while
OpenHands provides a platform for software-development agents that edit files,
execute commands, and browse~\cite{wang2024codeact,wang2025openhands}.  MemGPT
frames context management as virtual memory, whereas AIOS studies scheduling,
memory, storage, access control, and resource support for LLM agents%
~\cite{packer2023memgpt,mei2024aios}.

Self-evolving systems go beyond selecting from a fixed set of Tool calls.
Voyager accumulates skills, AFlow searches over code-represented workflows,
SkillWeaver synthesizes web skills, Alita constructs task-related capabilities,
and the Darwin G\"odel Machine modifies agent code subject to empirical
validation%
~\cite{wang2023voyager,zhang2024aflow,zheng2025skillweaver,qiu2025alita,zhang2025dgm}.
\sys is complementary: it neither selects an improvement nor establishes its
utility.  Instead, it provides a substrate for process lifecycle, authority,
information flow, durability, and evidence, allowing such mutable affordances
to run without treating their availability as a permission grant.

\subsection{Capabilities, policy, and delegation}

Classical capability systems bind designation to authority through protected
object references and support least authority and attenuation, with revocation
provided through system-specific mechanisms%
~\cite{dennis1966programming,miller2006robust,hardy1985keykos,shapiro1999eros,watson2010capsicum}.
Exokernels and library OSes place selected abstractions above a protected lower
interface~\cite{engler1995exokernel,madhavapeddy2013unikernels}.  Containers and
microVMs isolate host execution at a different layer%
~\cite{merkel2014docker,agache2020firecracker}.  \sys adopts the idea of a
stable lower boundary and an explicit authority vocabulary, but defines
agent-native subjects and resources: Tool tables, Skills, process images,
Objects, context, Human queues, provider endpoints, Task Authority ceilings, and
external-effect receipts.

Progent checks symbolic rules over tool names and arguments, automatically
applying verified policy narrowing while requiring approval for expansion%
~\cite{progent2025}.  AgentSpec provides customizable runtime enforcement for
agent actions, while Conseca associates contextual policies with task purpose%
~\cite{agentspec2026,conseca2025}.  Other recent work studies authenticated
delegation and statically tracked capabilities%
~\cite{authenticateddelegation2025,odersky2026tracked}.
Separately, SAGA governs inter-agent contact through user-defined policies and
provider-derived access-control tokens~\cite{saga2026}.

GNAP and the OAuth Security Best Current Practice address grant negotiation and
token security at network authorization boundaries~\cite{gnap2024,oauth2025}.
MCP standardizes
discovery, JSON-RPC messaging, standard transports, and optional authorization
for HTTP transports.  Its trust-and-safety guidance emphasizes user consent;
because the protocol cannot itself enforce consent or access control, Hosts
implement those mechanisms~\cite{mcp2026}.  These works target different
layers and can establish or constrain Host-side credentials and policy.  In
\sys, operation-time admission additionally intersects configured authority
with the invoking process's identity and Capabilities, any applicable Task
Authority ceiling, budgets, and primitive contracts; it separately requires
information-flow admission.  It then binds the resulting admission to durable
effect state and revalidates or reconciles the relevant bindings after a crash.
This dynamic check also reserves finite authority against a prepared external
effect, complementing static capture checks and inter-agent contact tokens%
~\cite{odersky2026tracked,saga2026}.

\subsection{Information flow and declassification}

Information-flow control (IFC) provides a vocabulary for confidentiality,
integrity, and end-to-end propagation%
~\cite{denning1976,myers1997dlm,myers1999jflow,sabelfeld2003}.  Work on
declassification classifies release policies by what information is released,
who may release it, where release occurs, and when it may occur%
~\cite{sabelfeld2009declassification}.  \sys uses conservative label joins, a
Host-owned recipient-trust registry, and an exact one-shot release bound to the
actor, Sink, payload, source versions, policy, and registry generation.  It
does not establish noninterference, automatically infer semantic independence,
or control a recipient after delivery.

Recent systems occupy different points in the agent-IFC design space.  Wu
et al.\ and CaMeL isolate planning or derive control flow from trusted input so
that, under their threat models, untrusted observations cannot redirect
execution; Fides formalizes planner-level label propagation and deterministic
security policies; permissive analysis instead identifies influential inputs
and regenerates outputs from a restricted context to derive less restrictive
labels~\cite{wu2024ifc,debenedetti2026camel,costa2025fides,siddiqui2025permissive}.
Several of these systems provide formal guarantees for specified planner
architectures, whereas permissive analysis emphasizes precision and utility.
\sys makes a different trade: it does not require a trusted plan and therefore
does not claim to prevent untrusted data from redirecting control flow.
Instead, it enforces authority and flow checks at provider and primitive
boundaries, integrates those checks with persistent processes and recovery
from ambiguous effects, and treats semantic classifiers as veto or escalation
evidence rather than as an allow oracle.

\subsection{Durable execution and external effects}

Sagas define compensating actions for long-lived transactions, while RIFL
suppresses duplicate RPC execution through stable request identities and
durable completion records~\cite{sagas1987,rifl2015}.  Durable Functions proves
that its compute--storage model simulates the fault-free high-level model and,
for deterministic orchestrations, that its replay model is bisimilar to the
compute--storage model.  The compute--storage result excludes external calls
because repeated work-item attempts can produce observable
duplicates~\cite{durablefunctions2021}.  For externally visible operations
routed through its API, Beldi provides crash-tolerant state equivalent to that
of a single failure-free stateful serverless function (SSF)
execution~\cite{beldi2020}.  ExoFlow
provides exactly-once DAG execution---visible outputs consistent with a
failure-free execution---under declared task semantics.  Externally visible
tasks must be idempotent or provide an idempotent rollback, while ExoFlow
coordinates checkpoints, rollback, and replay~\cite{exoflow2023}.  SagaLLM
applies saga ideas to LLM planning, while RAC adapts log-based compensation to
agent workflows and invokes compensating actions after
failures~\cite{sagallm2025,rac2026}.

Collectively, these systems illustrate that their guarantees for external
effects rely on explicit, scoped mechanisms---compensating transactions,
completion records atomically coupled to mutations, mediated operation APIs,
application-supplied idempotence or rollback semantics, or a proof whose scope
excludes external calls.  In these designs, a local transaction or replay log
alone is insufficient.

\sys does not claim general exactly-once execution or automatic compensation.
It records rollback class and reconciliation evidence but does not compensate
provider state; any compensation would itself be a separately authorized
Protected Operation.  Its narrower contribution couples durable intent and
receipt reconciliation to finite Capability and release reservations, Task
Authority, DataFlow, resource settlement, and an agent-process causal tree.  In
its recovery state machine, an unknown outcome blocks automatic replay until
registered reconciliation supplies sufficient evidence.

\subsection{Provenance and explainability}

W3C PROV defines a general entity/activity/agent provenance model, while LPM
and CamFlow capture whole-system provenance over kernel-mediated interactions
and information flows~\cite{w3cprov2013,lpm2015,camflow2017}.  PROV-AGENT
extends W3C PROV to agentic workflows, whereas AgentSight correlates LLM traffic
with kernel-observed effects for system-level observability and debugging%
~\cite{provagent2025,agentsight2025}.
AttriGuard uses counterfactual shadow replay to infer whether a tool invocation
is causally attributable to user intent rather than to untrusted
observations~\cite{attriguard2026}.  \sys makes no such semantic-causation
claim: its causal links encode durable, explicit structural relations among
Runtime records.  It also does not attempt whole-host provenance.  The
Explainable Operations view covers the recorded structural lineage from an LLM
call through Tool, syscall, primitive, decision, Human, provider, resource,
event, and audit records; a separate manifest records context materialization.
Its completeness predicate explicitly identifies required roles and missing
evidence and never treats an evidence graph as an authorization credential.

\subsection{Agent security evaluation}

Research on indirect prompt injection shows that untrusted observations can
elicit harmful actions~\cite{greshake2023not}.  InjecAgent, ToolEmu, and
AgentDojo provide benchmarks or environments for task-level security
evaluation~\cite{zhan2024injecagent,ruan2024toolemu,debenedetti2024agentdojo}.
SWE-bench, WebArena, and OSWorld measure useful behavior in software, web, and
desktop environments~\cite{jimenez2024swebench,zhou2024webarena,xie2024osworld}.
The \sys evaluation addresses a complementary question: whether authority,
flow, recovery, and evidence invariants hold as the Runtime action surface
changes.  It reports deterministic boundary tests separately from real-model
utility.  This paper neither reports AgentDojo results nor claims that its
synthetic live tasks replace broad task-success benchmarks.

\section{Limitations and Conclusion}
\label{sec:limits}

\subsection{Limitations and trusted boundary}

\textbf{Not a kernel sandbox.}  \sys runs inside a conventional Python process
and relies on the Host OS and providers.  Deno supervision, path checks, typed
Git command mediation, and transport bounds provide defense in depth.  Trusted
Runtime Modules and provider implementations, together with MCP stdio
executables and authorized shell or native programs, can exercise Host
authority outside the typed boundary.  Deployments requiring stronger
isolation should run such components in containers, WASM runtimes, or microVMs,
or move them behind remote-service boundaries.

\textbf{Prompt injection remains possible.}  An adversarial observation may
alter the model's plan, trigger repeated denied requests, consume budget, or
induce a harmful action that is nevertheless permitted by the granted task
policy.  The three-plane design constrains mediated authority and delivery; it
does not establish semantic correctness.  Semantic Shadow is disabled by
default, cannot provide a general allow predicate, and does not write classifier
findings back into ambient labels.

\textbf{Mediation has endpoints.}  Sink clearance applies only to deliveries
mediated by the Runtime.  Once a trusted shell, PTY, Human, LLM provider, MCP
server, JSON-RPC service, filesystem target, or Git remote receives data, its
subsequent handling, forwarding, and retention lie outside the threat model.
Labels provide neither encryption nor deletion guarantees.  The Runtime also
cannot observe effects produced by Host code that bypasses its primitives.

\textbf{Protocol and platform coverage is finite.}  The MCP implementation is
a governed client subset rather than an implementation of the full
specification.  Typed Git is limited to a configured local repository and its
existing remotes; it does not provide GitHub or GitLab product integration.
PTY and Deno containment remain subject to platform-specific release gates.
The local GUI is an operator console, not a hardened multi-tenant remote
control plane.  Each Task Run supervises a single process tree; Task Runs do
not form a distributed workflow scheduler.

\subsection{Conclusion}

Self-evolving LLM agents need a stable boundary beneath prompts, Tool schemas,
Skills, generated code, images, checkpoints, and remote-resource catalogs.  The
\sys design separates this boundary into three parts: conjunctive operation
admission, independent information-flow admission, and durable causal evidence
that never becomes authority.  A process's model-visible action surface may
evolve, but resource authority remains determined by Capabilities and any
applicable explicit or derived Task Authority ceiling when a concrete primitive
executes.  Data may leave only through a Host-resolved Sink for which the
clearance check passes; conditionally permitted high-sensitivity egress additionally requires
an exact one-shot release.

The implementation applies this principle through persistent processes, Object
Memory, Skill projection, syscall-mediated JIT Tools, typed providers, images,
checkpoints, Task Runs, and ordered startup recovery, subject to the
commit-bound limitations reported above.  Its Protected Operation protocol
makes provider ambiguity explicit: preparation is durably recorded before
dispatch, settlement links the available evidence, and reopening never implies
replay.  The resulting system does not prove that an agent will reason safely.
It provides a reference boundary that constrains and explains which requested
effects an evolving agent may make real.

\section*{Availability}
The implementation, documentation, benchmark harness and release
are available at \url{https://github.com/yingqi-z20/Agent-libOS}. 

\begin{acks}
Generative AI tools assisted with code development, manuscript organization,
and language polishing; they are not listed as authors.  The author reviewed
and revised all AI-assisted material and remains responsible for the technical
and implementation claims, references, and final manuscript.
\end{acks}

\bibliographystyle{plainnat}
\bibliography{references}

\end{document}